\documentclass[nojss]{jss}

\usepackage{thumbpdf,lmodern}

\usepackage{framed}

\usepackage{amsmath}
\usepackage{amsfonts}
\usepackage{amssymb}
\usepackage{bm}
\usepackage[usenames,dvipsnames]{xcolor}

\author{{\O}ystein S{\o}rensen\\University of Oslo
   \And Marta Crispino\\ Inria Grenoble
   \And Qinghua Liu\\University of Oslo
   \And Valeria Vitelli\\University of Oslo}
\Plainauthor{{{\O}}ystein S{{\o}}rensen, Second Author}

\title{\pkg{BayesMallows:} An \proglang{R} Package for the Bayesian Mallows Model}
\Plaintitle{BayesMallows: An R Package for the Bayesian Mallows Model}
\Shorttitle{\pkg{BayesMallows:} An \proglang{R} Package for the Bayesian Mallows Model}
\Abstract{
\pkg{BayesMallows} is an \proglang{R} package for analyzing data in the form of rankings or preferences with the Mallows rank model, and its finite mixture extension, in a Bayesian probabilistic framework. 

The Mallows model is a well-known model, grounded on the idea that the probability density of an observed ranking decreases exponentially fast as its distance to the location parameter increases. Despite the model being quite popular, this is the first Bayesian implementation that allows a wide choice of distances, and that works well with a large amount of items to be ranked. \pkg{BayesMallows} supports footrule, Spearman, Kendall, Cayley, Hamming and Ulam distances, allowing full use of the rich expressiveness of the Mallows model. This is possible thanks to the implementation of fast algorithms for approximating the partition function of the model under various distances. Although developed for being used in computing the posterior distribution of the model, these algorithms may be of interest in their own right. 

\pkg{BayesMallows} handles non-standard data: partial rankings and pairwise comparisons, even in cases including non-transitive preference patterns. The advantage of the Bayesian paradigm in this context comes from its ability to coherently quantify posterior uncertainties of estimates of any quantity of interest. These posteriors are fully available to the user, and the package comes with convienient tools for summarizing and visualizing the posterior distributions.
}

\Plainkeywords{rankings, permutations, Mallows model, Bayesian analysis, MCMC, importance sampling}

\Address{
  {{\O}}ystein S{{\o}}rensen\\
  Center for Lifespan Changes in Brain and Cognition\\
  Department of Psychology\\
  University of Oslo\\
  E-mail: \email{oystein.sorensen@psykologi.uio.no}
  
  Marta Crispino\\
  Univ. Grenoble Alpes, Inria, CNRS, LJK\\
  38000 Grenoble, France\\
  E-mail: \email{crispino.marta8@gmail.com}
  
  Qinghua Liu\\
  Department of Mathematics \\
  University of Oslo\\
  E-mail: \email{qinghual@math.uio.no}
  
  Valeria Vitelli\\
  Oslo Centre for Biostatistics and Epidemiology\\
  Department of Biostatistics\\
  University of Oslo\\
  E-mail: \email{valeria.vitelli@medisin.uio.no}
}

\begin{document}

\section[Introduction]{Introduction} \label{sec:intro}

Preference data are usually collected when individuals are asked to rank a set of items according to a certain preference criterion. The booming of internet-related activities and applications in recent years has led to a rapid increase of the amount of data that have ranks as their natural scale, however often in the form of partial or indirect rankings (pairwise preferences, ratings, clicks). The amount of readily available software handling preference data has consequently increased consistently in the last decade or so, but not many packages are flexible enough to handle all types of data mentioned above. The typical tasks when analyzing preference data are rank aggregation, classification or clustering, and prediction, where the latter task refers to the estimation of the individual rankings of the assessors (when completely or partially missing). These tasks can be addressed either via model-based inference or via heuristic machine learning algorithms, and (in the former case) with or without uncertainty quantification. However, not many methods allow combining various inferential tasks and individualized predictions.

The \proglang{R} \citep{R} package \pkg{BayesMallows} for analyzing ranking data, available at the Comprehensive \proglang{R} Archive Network (CRAN) at \url{https://CRAN.R-project.org/}, is one of the first attempts in this direction: \pkg{BayesMallows} implements full Bayesian inference for ranking data, and performs any of the tasks above in the framework of the Bayesian Mallows model (BMM) \citep{Mallows1957,Vitelli2018}. More specifically, \pkg{BayesMallows} allows for data in the forms of complete rankings, partial rankings, as well as possibly inconsistent pairwise comparisons. In these situations, it provides all Bayesian inferential tools for rank modeling with the BMM: it performs rank aggregation (estimation of a consensus ranking of the items), it can cluster together the assessors providing similar preferences (estimating both cluster specific model parameters, and individual cluster assignments, with uncertainty), it performs data augmentation for estimating the latent assessor-specific full ranking of the items in all missing data situations (partial rankings, pairwise preferences). The latter in particular, i.e., the possibility of predicting individual preferences for unranked items, enables the model to be used as a probabilistic recommender system. \pkg{BayesMallows} also enlarges the pool of distances that can be used in the Mallows model, and it supports some of the rank distances most used in the literature: Spearman's footrule (henceforth footrule), Spearman's rank correlation (henceforth Spearman), Cayley, Hamming, Kendall, and Ulam distances \citep[we refer to][for details on these]{Diaconis1988,Marden1995}. Finally, \pkg{BayesMallows} implements the IPFP algorithm for computing the partition function for the Mallows model (MM) \citep{Mukherjee2016} and the importance sampling algorithm described in \citet{Vitelli2018}. In addition to being used in the MM, these functions may be of interest in their own right.

Comparing with other available inferential software, we notice that not many packages allow for such flexibility, very few in combination with full Bayesian inference, and none when using the MM. Often machine learning approaches focus on either rank aggregation (i.e., consensus estimation), or individual rank prediction, while \pkg{BayesMallows} handles both. Since the BMM is fully Bayesian, all posterior quantities of interest are automatically available from \pkg{BayesMallows} for the first time for the MM. In addition, the package also has tools for visualizing posterior distributions, and hence, posterior quantities as well as their associated uncertainty. Uncertainty quantification is often critical in real applications - for recommender systems, the model should not spam the users with very uncertain recommendations. When performing subtype identification for cancer patients, a very uncertain cluster assignment might have serious consequences for the clinical practice, for example in treatment choice. The package also works well with a fairly large number of items, thanks to computational approximations and efficient \proglang{C++} programming. In conclusion, \pkg{BayesMallows} provides the first fully probabilistic inferential tool for the Mallows model with many different distances. It is flexible in the type of data it handles, and computationally efficient. We therefore think that this software tool will gain popularity, and prove its usefulness in many practical situations, many of which we probably cannot foresee now.

The paper is organized as follows. The BMM for ranking data is briefly reviewed in Section \ref{sec:background}, as well as its model extensions to different data types and to mixtures. Section \ref{sec:mcmc} includes details on the implementation of the inferential procedure. For a thorough discussion of both the model and its implementation we refer the interested reader to \citet{Vitelli2018}. An overview of existing \proglang{R} packages implementing the Mallows model (MM) is given in Section \ref{sec:packages}. The use of the \pkg{BayesMallows} package is presented, in the form of a quick-start guide, in Section \ref{sec:bayesmallows}. Section \ref{sec:concl} concludes the paper, also discussing model extensions that will come with new releases of the package.

\section{Background: The Bayesian Mallows Model for Rankings}\label{sec:background}

In this section we give the background for understanding the functions in the \pkg{BayesMallows} package. More details can be found in \citet{Vitelli2018} and \citet{Liu2019}. The section is organized as follows: we first clarify the notations that we will use throughout the paper (Section \ref{sec:notation}). We then briefly describe the BMM for complete ranking data (Section \ref{sec:model}), also focusing on the relevance of the choice of distance (Section \ref{sec:distances}). The last two sections focus on model extensions: partial and pairwise data (Section \ref{sec:partial}), non-transitive pairwise comparisons (Section \ref{sec:nt}), and mixtures (Section \ref{sec:clustering}). 

\subsection{Notations}\label{sec:notation}
Let us denote with $\mathcal{A}=\{A_1,...,A_n\}$ the finite set of labeled items to be ranked, and with $\mathcal{P}_n$ the space of $n$-dimensional permutations. A complete ranking of $n$ items is then a mapping $\bm{R}:\, \mathcal{A}\rightarrow \mathcal{P}_n$ that attributes a rank ${R}_i\in\{1,...,n\}$ to each item, according to some criterion. We here denote a generic complete ranking by $\bm{R}=(R_{1},...,R_{n})$, where $R_{i}$ is the rank assigned to item $A_i$. 

Note the intimate relation that exists between a ranking and pairwise preferences.
Given a pair of items $\{A_{i}, A_{k}\}$, we denote  a pairwise preference between them as $(A_{i}\prec A_{k})$, meaning that item $A_{i}$ is preferred to item $A_{k}$.
Given a full ranking $\bm{R}\in\mathcal{P}_n$, it is immediate to evince all the possible $n(n-1)/2$ pairwise preferences between the items taken in pairs, since the item in the pair having the lower rank is the preferred one:
\begin{equation*}\label{prefff}
\begin{split}(A_{t_1}\prec A_{t_2}) \text{     }\iff\text{     } R_{t_1}<{R}_{t_2},\quad\quad t_1,t_2=1,...,n,\quad t_1\not=t_2 ,
\end{split}\end{equation*}

Pairwise preference data are typically incomplete, meaning that not all the $n(n-1)/2$ pairwise preferences that determine each individual ranking are present, or they can contain non-transitive patterns, that is, one or more pairwise preferences contradict what is implied by other pairwise preferences. In this package we deal with partial and possibly non-transitive pairwise preferences. 

\subsection{The BMM for Complete Rankings}\label{sec:model}
The MM for ranking data \citep{Mallows1957} specifies  the probability density for a ranking $\mathbf{r}\in\mathcal{P}_n$ as follows
\begin{equation} \label{eq:mallows_distribution}
P(\mathbf{r} | \alpha, \bm{\rho}) = \frac{1}{Z_{n}\left(\alpha\right)} \exp\left[-\frac{\alpha}{n}d\left(\mathbf{r}, \bm{\rho}\right)\right] 1_{\mathcal{P}_{n}}(\mathbf{r})
\end{equation}
where $\bm{\rho} \in \mathcal{P}_{n}$ is the location parameter representing the consensus ranking, $\alpha\geq0$ is the scale parameter (precision), $Z_{n}(\alpha)$ is the normalizing constant (or partition function), $d(\cdot, \cdot)$ is a right-invariant distance among rankings \citep{Diaconis1988}, and $1_{S}(\cdot)$ is the indicator function for the set $S$. 

In the complete data case, $N$ assessors have provided complete rankings of the $n$ items in $\mathcal{A}$ according to some criterion, yielding the permutation $\mathbf{R}_{j} = (R_{1j}, \dots, R_{nj})$ for assessor $j,$ $j=1,\ldots,N$. The likelihood of the $N$ observed rankings $\mathbf{R}_{1}, \dots, \mathbf{R}_{N}$, assumed conditionally independent given $\alpha$ and $\bm{\rho}$, is 
\begin{equation} \label{eq:mallows_likelihood}
P\left(\mathbf{R}_{1}, \dots, \mathbf{R}_{N} | \alpha, \bm{\rho}\right) = \frac{1}{Z_{n}\left(\alpha\right)^N} \exp\left[-\frac{\alpha}{n}\sum_{j=1}^Nd\left(\mathbf{R}_j, \bm{\rho}\right)\right] \prod_{j=1}^N 1_{\mathcal{P}_n}(\mathbf{R}_j).
\end{equation}

According to the BMM introduced in \citet{Vitelli2018}, prior distributions have to be elicited on every parameter of interest. A truncated exponential prior distribution was specified for $\alpha$
\begin{equation}
\pi\left(\alpha | \lambda\right) = \frac{\lambda \exp\left(-\lambda \alpha\right) 1_{\left[0, \alpha_{\text{max}}\right]}\left(\alpha\right)}{1 - \exp(-\lambda \alpha_{\text{max}})},
\end{equation}
where $\lambda$ is a rate parameter, small enough to ensure good prior dispersion, and $\alpha_{\text{max}}$ is a cutoff, large enough to cover reasonable $\alpha$ values. A uniform prior $\pi(\bm{\rho})$ on $\mathcal{P}_{n}$ was assumed for $\bm{\rho}$. It follows that the posterior distribution for $\alpha$ and $\bm{\rho}$ is
\begin{equation} \label{eq:mallows_posterior}
P\left(\alpha, \bm{\rho} | \mathbf{R}_{1}, \dots, \mathbf{R}_{N}\right) \propto \frac{1_{\mathcal{P}_{n}}(\bm{\rho})}{Z_{n}(\alpha)^{N}} \exp\left[-\frac{\alpha}{n}\sum_{j=1}^{N}d(\mathbf{R}_{j}, \bm{\rho}) - \lambda \alpha\right] \frac{ 1_{\left[0, \alpha_{\text{max}}\right]}\left(\alpha\right)}{1 - \exp(-\lambda \alpha_{\text{max}})} .
\end{equation}

Inference on the model parameters is based on a Metropolis-Hastings (M-H) Markov Chain Monte Carlo (MCMC) algorithm, described in  \citet{Vitelli2018}. Some details relevant for a correct use of this package are also given in Section \ref{sec:mainalgo}.

\subsection{Distance Measures and Partition Function}\label{sec:distances}

The reason why the partition function $Z_{n}\left(\alpha\right)$ reported in \eqref{eq:mallows_distribution}, \eqref{eq:mallows_likelihood} and \eqref{eq:mallows_posterior} does not depend on the latent consensus ranking $\bm{\rho}$ is that the chosen distance $d(\cdot, \cdot)$ is right-invariant. A right-invariant distance \citep{Diaconis1988} is unaffected by a relabelling of the items. Right-invariant distances  play an important role in the MM, and for this reason the \pkg{BayesMallows} package only handles right-invariant distances. 

The choice of distance affects the model fit to the data and the results of the analysis, and is crucial also because of its role in the partition function computation. Some right-invariant distances allow for analytical computation of the partition function, and for this reason they have become quite popular. In particular, the MM with Kendall \citep{Meila2010,Lu2015},  Hamming \citep{irurozki2014Ham} and  Cayley \citep{irurozki2016sampling} distances have a closed form of $Z_n(\alpha)$ due to \citet{Fligner1986}. There are however important and natural right-invariant distances for which the computation of the partition function is NP-hard, in particular the footrule (${l}_1$) and the Spearman (${l}_2$) distances. For precise definitions of all distances involved in the MM we refer to \citet{Marden1995}.

The \pkg{BayesMallows} package handles the footrule, Spearman, Cayley, Hamming, Kendall, and Ulam distances. In the case of Cayley, Hamming, Kendall, and Ulam the exact partition function is implemented. In the case of footrule and Spearman there are three possibilities supported by the package: (i) for moderately large values of $n$, an exact implementation is available, and the maximum value of $n$ for which this is available depends on the chosen distance; (ii) for larger values of $n$, an importance sampling approximation scheme is proposed, and the approximated $Z_{n}\left(\alpha\right)$ has to be computed off-line over a grid of $\alpha$ values; (iii) for very large  $n$, the asymptotic approximation of \citet{Mukherjee2016} is implemented, which also requires off-line computation.

Further details on the different strategies for computing the partition function are given in Section \ref{sec:normconst}. A thorough description of these themes can also be found in \citet{Vitelli2018}.

\subsection{Partial Rankings and Transitive Pairwise Comparisons}\label{sec:partial}

When complete rankings of all items are not readily available, the BMM can still be used by applying data augmentation techniques. Partial rankings can occur because ranks are missing at random, because the assessors have only ranked their top-$k$ items, or because they have been presented with a subset of items. In more complex situations, data do not include ranks at all, but the assessors have only compared pairs of items and given a preference between the two. The Bayesian data augmentation scheme can still be used to handle such pairwise comparison data, thus providing a solution which is fully integrated into the Bayesian inferential framework.

Let us start by considering the case of top-$k$ rankings. Suppose that each assessor $j$ has chosen a set of preferred items $\mathcal{A}_{j} \subseteq \mathcal{A}$, which were given ranks from $1$ to $n_j=|\mathcal{A}_{j}|$. Now $R_{ij} \in \{1,\ldots,n_j\}$ if $A_{i} \in \mathcal{A}_{j}$, while for $A_i \in \mathcal{A}^c_{j}$, $R_{ij}$ is unknown, except for the constraint $R_{ij} > n_j$, $j=1,\dots,N.$ The augmented data vectors $\tilde{\mathbf{R}}_{1},\dots,\tilde{\mathbf{R}}_{N}$ are introduced in the model to include the missing ranks, which are estimated as latent parameters. Let $\mathcal{S}_{j} = \{ \tilde{\mathbf{R}}_{j} \in \mathcal{P}_{n} : \tilde{R}_{ij} = R_{ij} \text{ if } A_{i} \in \mathcal{A}_{j} \}, ~ j=1,\dots,N$ be the set of possible augmented random vectors, including the ranks of the observed top-$n_j$ items together with the unobserved ranks, which are assigned a uniform prior on the permutations of $\{n_j+1,\ldots,n\}$.
The goal is to sample from the posterior distribution
\begin{align}\label{eq:mallows_partial}
P\left(\alpha, \bm{\rho}| \mathbf{R}_{1},\dots, \mathbf{R}_{N}\right) = \sum_{\tilde{\mathbf{R}}_{1} \in \mathcal{S}_{1}} \dots \sum_{\tilde{\mathbf{R}}_{N} \in \mathcal{S}_{N}} P\left( \alpha, \bm{\rho} , \tilde{\mathbf{R}}_{1},\dots, \tilde{\mathbf{R}}_{N}|\mathbf{R}_{1},\dots, \mathbf{R}_{N}\right).
\end{align}
The augmentation scheme amounts to alternating between sampling $\alpha$ and $\bm{\rho}$ given the current values of the augmented ranks using the posterior given in (\ref{eq:mallows_posterior}), and sampling the augmented ranks given the current values of $\alpha$ and $\bm{\rho}$. For the latter task, once $\alpha$, $\bm{\rho}$ and the observed ranks $\mathbf{R}_{1},\dots, \mathbf{R}_{N}$ are fixed, one can see that $\tilde{\mathbf{R}}_{1},\dots, \tilde{\mathbf{R}}_{N}$ are conditionally independent, and that each $\tilde{\mathbf{R}}_{j}$ only depends on the corresponding ${\mathbf{R}}_{j}$. As a consequence, the updating of new augmented vectors $\tilde{\mathbf{R}}^\prime_{j}$ is performed independently, for each $j =1,\dots,N$. 

The above procedure can also handle more general situations where missing rankings are not necessarily the bottom ones, and where each assessor is asked to provide the mutual ranking of some possibly random subset $\mathcal{A}_j\subseteq\mathcal{A}$ consisting of $n_j\leq n$ items. Note that the only difference from the previous formulation is that each latent rank vector $\tilde{\mathbf{R}}_j$ takes values in the set  $\mathcal{S}_j=\{\tilde{\mathbf{R}}_j\in\mathcal{P}_n:(R_{i_1j}<R_{i_2j}) \land (A_{i_1},A_{i_2}\in \mathcal{A}_j)\Rightarrow \tilde{R}_{i_1j}<\tilde{R}_{i_2j}\}$. Also in this case the prior for $\tilde{\mathbf{R}}_j$ is assumed uniform on $\mathcal{S}_j$.

In the case of pairwise comparison data, let us call $\mathcal{B}_{j}$ the set of all pairwise preferences stated by assessor $j$, and let 
$\mathcal{A}_{j}$ be the set of items appearing at least once in $\mathcal{B}_{j}$.
Note that the items in $\mathcal{A}_{j}$ are not necessarily fixed to a given rank but may only be given some partial ordering. For the time being, we assume that the observed pairwise orderings in $\mathcal{B}_{j}$ are transitive, i.e., mutually compatible, and define by $\text{tc}(\mathcal{B}_{j})$ the transitive closure of $\mathcal{B}_{j}$, containing all pairwise orderings of the elements in $\mathcal{A}_{j}$ induced by $\mathcal{B}_{j}$. The model formulation remains the same as in the case of partial rankings, with the prior for the augmented data vectors $\tilde{\mathbf{R}}_{1},\dots,\tilde{\mathbf{R}}_{N}$ being uniform on the set $\mathcal{S}_j$ of rankings that are compatible with the observed data.

More details on the algorithms used to handle the situations described in this section are provided in Section \ref{sec:mainalgo}. 

\subsection{Non-Transitive Pairwise Comparisons}\label{sec:nt}

It can happen in real applications that individual pairwise comparison data are non-transitive, that is, they may contain a pattern of the form $x \prec y\,,\,\, y \prec z $ but $z \prec x $. This is typically the case of data collected from internet user activities, when the pool of items is very large: non-transitive patterns can arise for instance due to assessors' inattentiveness, uncertainty in their preferences, and actual confusion, even when one specific criterion for ranking is used. Another frequent situation is preferences collected over time, so that assessors might change opinion along data collection. This setting is considered in \citet{crispino2017bayesian}, where the model for transitive pairwise comparisons of Section \ref{sec:partial} is generalized to handle situations where non-transitivities in the data occur. Note that the kind of non-transitivity that is considered in \citet{crispino2017bayesian} considers only the individual level preferences. A different type of non-transitivity, which we do not consider here, arises when aggregating preferences across assessors, as under Condorcet \citep{Condorcet1785} or Borda \citep{Borda1781} voting rules. 

The key ingredient of this generalization consists of adding one layer of latent variables to the model hierarchy, accounting for the fact that assessors can make mistakes. The main assumption is that the assessor makes pairwise comparisons based on her latent full rankings $\tilde{\bm R}$. A mistake is defined as an inconsistency between one of the assessor's pairwise comparisons and $\tilde{\bm R}$. Suppose each assessor $j=1,\dots,N$ has assessed $M_j$ pairwise comparisons, collected in the set $\mathcal{B}_j,$ and assume the existence of latent ranking vectors $\tilde{\bm R_j}$, $j=1,\dots,N$. Differently from Section \ref{sec:partial}, since $\mathcal{B}_j$ is allowed to contain non-transitive pairwise preferences, the transitive closure of $\mathcal{B}_j$ is not defined, and the posterior density \eqref{eq:mallows_partial} cannot be evaluated. 
In this case, the posterior takes the form,
\begin{equation}\label{incons}\begin{split}
P(\alpha,\bm{\rho}| \mathcal{B}_{1},...,\mathcal{B}_{N})=&\sum_{\tilde{\bm{R}}_1\in\mathcal{P}_n}\dots \sum_{\tilde{\bm{R}}_N\in\mathcal{P}_n} P(\alpha, \bm{\rho},\tilde{\bm{R}}_1,...,\tilde{\bm{R}}_N|\mathcal{B}_{1},...,\mathcal{B}_{N})=\\
=&\sum_{\tilde{\bm{R}}_1\in\mathcal{P}_n}\dots \sum_{\tilde{\bm{R}}_N\in\mathcal{P}_n}
P(\alpha, \bm{\rho}|\tilde{\bm{R}}_1,...,\tilde{\bm{R}}_N)P(\tilde{\bm{R}}_1,...,\tilde{\bm{R}}_N|\mathcal{B}_{1},...,\mathcal{B}_{N})\\
\end{split}\end{equation}
where the term $P(\tilde{\bm{R}}_1,...,\tilde{\bm{R}}_N|\mathcal{B}_{1},...,\mathcal{B}_{N})$ models the presence of mistakes in the data, while in the case of transitive pair comparisons it was implicitly assumed equal to 1 if each augmented ranking $\tilde{\bm R}_j$ was compatible with the partial information contained in $\mathcal{B}_j$, and 0 otherwise. 

Two models for \eqref{incons} are considered in \citet{crispino2017bayesian}: the Bernoulli model, which accounts for random mistakes, and the Logistic model, which lets the probability of making a mistake depend on the similarity of the items being compared. The Bernoulli model states that:
\begin{equation}\label{eq:bm}P(\tilde{\bm{R}}_1,...,\tilde{\bm{R}}_N|\theta, \mathcal{B}_{1},...,\mathcal{B}_{N})\propto \theta^M(1-\theta)^{\sum_j M_j-M},\quad\theta\in[0,0.5)
\end{equation}
where $M$ counts the number of times the observed preferences contradict what is implied by the ranking $\tilde{\bm{R}}_j$, and the parameter $\theta$ is the probability of making a mistake in a single pairwise preference. $\theta$ is \emph{a priori} assigned a truncated Beta distribution on the interval $[0,0.5)$ with given hyper-parameters $\kappa_1$ and $\kappa_2$, conjugate to the Bernoulli model \eqref{eq:bm}. The Logistic model is a generalization of \eqref{eq:bm} where, instead of assigning a constant value $\theta$ to the probability of making a mistake, it depends on the distance between the ranks of the two items under comparison. In \citet{crispino2017bayesian} the Logistic model gave results very similar to the Bernoulli model, and currently only the Bernoulli model is available in \pkg{BayesMallows}. The sampling scheme is similar to the one used for the case of transitive pairwise preferences, apart from an additional step for updating $\theta$, and the augmentation scheme for $\tilde{\bm{R}}_j$, which is slightly different. We refer to \citet{crispino2017bayesian} for details.

\subsection{Clustering}\label{sec:clustering}

The assumption, implicit in the discussion so far, that there exists a unique consensus ranking shared by all assessors is unrealistic in most real applications: the BMM thus handles the case in which the rankings can be modeled as a sample from a finite mixture of MMs. In the following brief discussion we assume that the data consist of complete rankings, but \pkg{BayesMallows} can fit a mixture based on any kinds of preference data described so far.

Let $z_1,\ldots,z_N \in \{1,\ldots,C\}$ assign each assessor to one of $C$ clusters, and let the rankings within each cluster $c \in \{1,\ldots,C\}$ be described by an MM with parameters $\alpha_c$ and $\bm{\rho}_c$. The likelihood of the observed rankings $\mathbf{R}_1,\ldots,\mathbf{R}_N$ is given by
\begin{align} \label{eq:mallows_mixture}
P\left(\mathbf{R}_{1},\dots,\mathbf{R}_{N} \Big| \left\{\bm{\rho}_{c},\alpha_{c}\right\}_{c=1,...,C}, \{z_j\}_{j=1,\ldots,N} \right) = \prod_{j=1}^N \frac{1_{\mathcal{P}_n}(\mathbf{R}_{j})}{Z_{n}(\alpha_{z_j})}\exp\left[-\frac{\alpha_{z_j}}{n} d(\mathbf{R}_{j},\bm{\rho}_{z_j}) \right],
\end{align}
where conditional independence is assumed across the clusters. We also assume independent truncated exponential priors for the scale parameters and independent uniform priors for the consensus rankings. The cluster labels $z_1,\ldots,z_N$ are a priori assumed conditionally independent given the clusters mixing parameters $\tau_1,...,\tau_C$, and are assigned a uniform multinomial. Finally $\tau_1,\ldots,\tau_C$ (with $\tau_c\geq0, \ c=1,\ldots,C$ and $\sum_{c=1}^C \tau_c = 1$) are assigned the standard symmetric Dirichlet prior of parameter $\Psi$, thus implying a conjugate scheme. The posterior density is then given by
\begin{equation}\label{eq:post_mixture}
P\left(\{\bm{\rho}_{c},\alpha_{c}, \tau_c\}_{c=1}^C, \{z_j\}_{j=1}^N \Big|\mathbf{R}_{1},\dots,\mathbf{R}_{N}\right)
\propto \left[\prod_{c=1}^Ce^{-\lambda\alpha_c}\tau_c^{\Psi-1}\right]\left[
\prod_{j=1}^N\frac{\tau_{z_j}e^{-\frac{\alpha_{z_j}}{n} d(\mathbf{R}_{j},\bm{\rho}_{z_j})}}{Z_{n}(\alpha_{z_j})}\right].
\end{equation}


Some details on the algorithm used to handle the mixture model extension will be provided in Section \ref{sec:mainalgo}. 

\section{Learning the Bayesian Mallows Model}\label{sec:mcmc}

In this section we briefly give some additional details regarding the implementation of the models described in Section \ref{sec:background}. The BMM implementation is thoroughly described in \citet{Vitelli2018}.

\subsection{Details on the MCMC Procedures}\label{sec:mainalgo}
In order to obtain samples from the posterior density of equation \eqref{eq:mallows_posterior}, the \pkg{BayesMallows} package implements an MCMC scheme iterating between (i) updating $\bm\rho$ and (ii) updating $\alpha$ \citep[Algorithm 1 of][]{Vitelli2018}. The leap-and-shift proposal distribution, which is basically a random local perturbation of width $L$ of a given ranking, is used for updating $\bm\rho$ in the (i) step. The $L$ parameter of the leap-and-shift proposal controls how far the proposed ranking is from the current one, and it is therefore linked to the acceptance rate. The recommendation given in \citet{Vitelli2018} is to set it to $L=n/5$, which is also the default value in \pkg{BayesMallows}, but the user is allowed to choose a different value. For updating $\alpha$ in step (ii), a log-normal density is used as proposal, and its variance $\sigma_{\alpha}^{2}$ can be tuned to obtain a desired acceptance rate. 

As mentioned in Section \ref{sec:partial}, the MCMC procedure for sampling from the posterior densities corresponding to the partial data cases \citep[Algorithm 3 of][]{Vitelli2018} has an additional M-H step to account for the update of the augmented data rankings $\{\tilde{\mathbf{R}}_j\}_{j=1}^N$.  
In the case of partial rankings, for updating the augmented data $\tilde{\mathbf{R}}_j$, $j=1,...,N$ we use a uniform proposal on the set of rankings compatible with the partial data, $\mathcal{S}_j$. In the case of pairwise preferences, due to the increased sparsity in the data, we instead implemented a modified parameter-free leap-and-shift proposal distribution, which proposes a new augmented ranking by locally permuting the ranks in $\tilde{\mathbf{R}}_j$ within the constraints given by $\mathcal{B}_j$ \citep[Section 4.2]{Vitelli2018}.

The generalization to non-transitive pairwise comparisons, outlined in Section 4 of \citet{crispino2017bayesian}, requires further considerations. First, in the M-H step for updating the augmented data rankings, the modified parameter-free leap-and-shift proposal has to be replaced by a Swap proposal, whose tuning parameter $L^\star$ is the maximum allowed distance between the ranks of the swapped items. Second, the model for mistakes makes it necessary to add a completely new step in the procedure: a Gibbs step in the case of the Bernoulli model for the update of $\theta$, and two M-H steps, to update $\beta_0$ and $\beta_1$, in the case of the Logistic model. In the latter case, both values are proposed by sampling from a log-normal density, with variance to be fixed depending on the desired acceptance rate \citep[see also][Supplementary material]{crispino2017bayesian}. 

The MCMC algorithm for sampling from the mixture model posterior \eqref{eq:post_mixture} \citep[Algorithm 2 of][]{Vitelli2018} alternates between updating $\{\bm{\rho}_c, \alpha_c\}_{c=1}^C$ in a M-H step, and $\{\tau_c, z_j\}_{c=1, j=1}^{C, N}$ in a Gibbs sampling step, in addition to the necessary M-H steps for data augmentation or estimation of error models, as outlined above.

\subsection{Partition Function}\label{sec:normconst}
As already mentioned in Section \ref{sec:distances}, when the distance in the BMM is chosen to be the footrule or the Spearman, the partition function $Z_n(\cdot)$ does not have a closed form. Hence, in these situations \pkg{BayesMallows} allows for three different choices, which the user may employ depending on the value of $n$: (a) exact calculation, (b) Importance Sampling (IS), and (c) the asymptotic approximation due to \citet{Mukherjee2016}.

The package contains integer sequences for the exact calculation of the partition function with footrule distance for up to $n=50$ items, and with the Spearman distance for up to $n=14$ items \citep[see][Section 2.1]{Vitelli2018}. 
These sequences are downloaded from the On-Line Encyclopedia of Integer Sequences \citep{Sloane2017}. 

The IS procedure can be used to compute an off-line approximation $\hat{Z}_n(\alpha)$ of $Z_n(\alpha)$ for the specific value of $n$ which is needed in the application at hand. The IS estimate $\hat{Z}_n(\alpha)$ is computed on a grid of $\alpha$ values provided by the user, and then a smooth estimate obtained via a polynomial fit is returned to the user, who can also select the degree of the polynomial function. Finally, the user should set the number $K$ of IS iterations, and we refer to \citet{Vitelli2018} for guidelines on how to select a large enough value for $K$. Note that the procedure might take a long time, depending on $K$, $n$, and on how the grid for $\alpha$ is specified (in our experience, values of $n$ larger than approximately 100 might require $K$ to be as large as $10^8$ in order for the IS to provide a good estimate, and hence a quite long computing time).

The Iterative Proportional Fitting Procedure (IPFP) of \citet[][Theorem 1.8]{Mukherjee2016} is also implemented, yielding a numeric evaluation of $Z_\text{lim}(\cdot)$, the asymptotic approximation to $Z_n(\cdot)$. In this case the user needs to specify two parameters: the number of iterations $m$ to use in the IPFP, and the dimension $K$ of the grid approximating the continuous domain where the limit is computed. Values of $m$ and $K$ have been suggested by \citet{Mukherjee2016}, and we refer to the Supplementary Material of \citet{Vitelli2018} for more details.

\subsection{Sampling from the Bayesian Mallows Model}\label{sec:sampling}
When one needs to sample from the MM with Cayley, Hamming, Kendall, or Ulam, we suggest to use the \pkg{PerMallows} package, which is optimized for this task. We instead provide a procedure for sampling from the MM with footrule and Spearman.

The procedure to generate a random sample of size $N$ from the MM is straightforward, and described in Appendix C of \citet[][]{Vitelli2018}. Basically, we run the MCMC algorithm for the BMM with complete data, with both the consensus ranking $\bm{\rho}\in\mathcal{P}_n$, and $\alpha\geq0$ being fixed, and with a given distance measure $d(\cdot,\cdot)$, until convergence. We then take $N$ rankings with a large enough interval between each of them to achieve independence. 

\section{Packages Implementing the Mallows Model}\label{sec:packages}
This section gives an overview of the existing \proglang{R} packages for fitting the MM. We refer to the supplementary online material for \proglang{R} scripts illustrating the use of these packages and tests supporting our conclusions.

\begin{itemize}
\item \pkg{PerMallows} \citep{Irurozki2016} is the package that comes closest in functionality to \pkg{BayesMallows}. It contains functions for learning and sampling from the frequentist versions of the MM and generalized Mallows model (GMM) \citep{Fligner1986}. Compared to \pkg{BayesMallows}, it lacks support for footrule or Spearman distance, and it is not a Bayesian method, so it does not compute uncertainty ranges for the estimated parameters. In addition \pkg{PerMallows} handles only complete rankings, and does not provide functionality for computation of mixture models. According to \citet[Table 1]{Irurozki2016}, computing the maximum likelihood estimates (MLE) of $\alpha$ and $\rho$ using the function \code{lmm} is possible when $n < 80$ for Kendall, $n < 250$ for Cayley, $n < 90$ for Hamming and $n < 100$ for Ulam. Our experiments suggest that these estimates are conservative, and that even larger numbers of items are fit rapidly. Hence, \pkg{PerMallows} seems to be a good choice for modeling with complete data without clusters, when the supported distance measures are appropriate and uncertainty estimates are not sought. \pkg{PerMallows} also has very efficient functions for sampling from the MM with Cayley, Hamming, Kendall, and Ulam distances. 

\item \pkg{pmr} \citep{Lee2013} provides summary statistics, visualization, and model fitting tools for complete ranking data in the MM, as well as other models. The function \code{dbm} returns the MLE of $\alpha$ together with its variance. The MLE of $\rho$, however, is not returned, but printed to the console, and no uncertainty estimates are given. Internally, \code{dbm} generates a matrix of sixe $n! \times n$ containing all possible permutations of the $n$ items. As a result, it quickly runs into memory issues. In our tests, \code{pmr} was not able to handle a ranking dataset with $n = 10$ items.

\item \pkg{rankdist} \citep{rankdist} implements distance-based probability models for ranking data as described in \citet{Alvo2014}, returning MLEs for $\alpha$ and $\rho$, but no uncertainty estimates. The package handles a large number of distances and supports mixture models, but in our experiments a warning was issued when using mixtures with all distances except Kendall. \pkg{rankdist} also implements the GMM \citep{Fligner1986}. However, for Cayley, footrule, Hamming, and Spearman distances, it generates an $n! \times n$ matrix internally, causing our \proglang{R} session to crash with $n \geq 10$ items, hence limiting its applicability. For Kendall, on the other hand, \pkg{rankdist} appears to work fine both with a large number of items, and with mixtures.
\end{itemize}

As will be illustrated in the next Section, \pkg{BayesMallows} provides many new functionalities not implemented in these packages.

\section{The BayesMallows Package}\label{sec:bayesmallows}

The \pkg{BayesMallows} package runs on all platforms supported by CRAN. It can be installed with the command:
\begin{Schunk}
\begin{Sinput}
R> install.packages("BayesMallows")
\end{Sinput}
\end{Schunk}
The package is loaded into the current \proglang{R} session as follows:
\begin{Schunk}
\begin{Sinput}
R> library("BayesMallows")
\end{Sinput}
\end{Schunk}
An overview of the most used functions is provided in Table~\ref{tab:functions}, and Table~\ref{tab:data} lists example datasets included in the package. The main function of \pkg{BayesMallows} is \code{compute_mallows}, which computes the posterior distribution of the BMM. Its arguments are listed in Table~\ref{tab:arguments}. The following sections demonstrate \pkg{BayesMallows} with example use cases.

\begin{table}[t!]
\centering
\begin{tabular}{lp{9cm}}
\hline
Function Name                   & Description \\ \hline
\code{assess_convergence}  & Trace plots for studying convergence of the M-H algorithm. \\ 
\code{assign_cluster} & Compute the cluster assignment (hard or soft) for each assessor. \\
\code{compute_consensus}  & Rank the items according to their cumulative probability (CP) or maximum \emph{a posteriori} (MAP) consensus ranking. \\
\code{compute_mallows}          & Compute the posterior distribution of the BMM. This is the main function of the package. Returns an object of \code{S3} class \code{BayesMallows}. The arguments to \code{compute_mallows} are listed in Table~\ref{tab:arguments}. \\
\code{compute_mallows_mixtures}  & Compute multiple Mallows models with different number of mixture components. Returns a \code{list} of \code{S3} class \code{BayesMallowsMixtures}, containing the fitted models. This function is typically used in combination with \code{plot_elbow}, and can be run in parallel. \\ 
\code{compute_posterior_intervals} & Compute Bayesian posterior intervals for the parameters of the BMM. \\
\code{estimate_partition_function} & Estimate the partition function of the BMM for a given distance measures, using either importance sampling or the IPFP algorithm. Can be run in parallel. \\
\code{generate_initial_ranking} & Generate a rank matrix consistent with the transitive closure of a set of pairwise preferences. Can be run in parallel. \\
\code{generate_transitive_closure} & Generate the transitive closure of a set of pairwise preferences. Can be run in parallel. \\
\code{plot_elbow} & Create an elbow plot for comparing models with different number of mixtures. \\
\code{plot.BayesMallows}  & \code{S3} method for plotting posterior distributions of the model parameters. \\ 
\code{sample_mallows} & Obtain random samples from the Mallows model, using footrule, Spearman, Cayley, Hamming, Kendall, or Ulam distance. \\ \hline
\end{tabular}
\caption{\label{tab:functions} Overview of the most used functions in \pkg{BayesMallows}, in alphabetic order.}
\end{table}

\begin{table}[t!]
\centering
\begin{tabular}{lp{9cm}}
Name & Description \\ \hline
\code{beach_preferences} & Stated pairwise preferences between random subsets of 15 images of beaches, by 60 assessors \citep[Section 6.2]{Vitelli2018}. \\
\code{sushi_rankings} & Complete rankings of 10 types of sushi by 5000 assessors \citep{Kamishima2003}. \\
\code{potato_visual} & Complete rankings of 20 potatoes by weight, based on visual inspection, by 12 assessors \citep{Liu2019}. \\
\code{potato_weighing} & Complete rankings of 20 potatoes by weight, where the assessors were allowed to weigh the potatoes in their hands, by 12 assessors \citep{Liu2019}. \\
\code{potato_true_ranking} & Vector of true weight rankings for the 20 potatoes in the example datasets \code{potato_visual} and \code{potato_weighing} \citep{Liu2019}.\\
\hline
\end{tabular}
\caption{\label{tab:data} Example datasets in \pkg{BayesMallows}.}
\end{table}

\begin{table}[t!]
\centering
\begin{tabular}{lp{9cm}}
 \hline
 Argument & Description (default values in parentheses)    \\ \hline
 \code{rankings}           & A \code{matrix} of ranked items. Required if \code{preferences = NULL}. (\code{NULL})\\
 \code{preferences} & A \code{data.frame} of pairwise comparisons. Required if \code{rankings = NULL}. (\code{NULL})\\
\code{metric} & Distance measure to use. (\code{"footrule"})  \\
\code{error_model} & One of \code{NULL} and \code{"bernoulli"}. (\code{NULL})\\
\code{n_clusters} & Number of mixture components. (\code{1L})  \\
\code{save_clus} & Whether to save cluster assignments during MCMC. (\code{FALSE}) \\
 \code{clus_thin} & Thinning to use when saving cluster assignments. (\code{1L})  \\
\code{nmc} & Number of MCMC iterations. (\code{2000L}) \\
\code{leap_size} & Step size of leap-and-shift proposal for $\bm{\rho}$. (\code{max(1L, floor(n_items/5))}) \\
\code{swap_leap} & Step size of swap proposal for non-transitive pairwise comparisons. (1L) \\
\code{rho_init} & Optional initial value of $\bm{\rho}$. (\code{NULL}) \\
\code{rho_thinning} & Thinning of $\bm{\rho}$. (\code{1L}) \\
\code{alpha_prop_sd} & Standard deviation of the lognormal proposal distribution used for $\alpha$. (\code{0.1}) \\
\code{alpha_init} & Initial value of $\alpha$. (\code{1}) \\
\code{alpha_jump} & How many times to sample $\bm{\rho}$ between each sampling of $\alpha$. (\code{1L}) \\
\code{lambda} & Rate parameter of exponential prior for $\alpha$. (\code{0.1}) \\
\code{alpha_max} & Truncation parameter in the prior for $\alpha$. (\code{1e6}) \\
\code{psi} & Concentration parameter $\Psi$ of Dirichlet prior for cluster probabilities. (\code{10L}) \\
\code{include_wcd} & Whether to save within-cluster distances in MCMC. (\code{n_clusters > 1}) \\
\code{save_aug} & Whether to save augmentated data during MCMC. (\code{FALSE}) \\
\code{aug_thinning} & Thinning to use for augmented data. (\code{1L}) \\
\code{logz_estimate} & Estimate of $\log\{Z_{n}(\alpha)\}$. (\code{NULL})\\
\code{verbose} & Whether to print information on progress. (\code{FALSE})\\
\code{validate_rankings} & Whether to check that \code{rankings} contains proper permutations. (\code{TRUE}) \\
\code{constraints} & Constraint set used internally when augmenting ranks. Precomputing and providing it as an argument may save time. (\code{NULL})\\
\code{save_ind_clus} & Whether to save individual cluster probabilities during MCMC. (\code{FALSE})\\
\code{seed} & Random number seed. (\code{NULL})\\
\hline
\end{tabular}
\caption{\label{tab:arguments} Arguments to function \code{compute\_mallows}. Arguments are optional unless otherwise stated.}
\end{table}

\subsection{Analysis of Complete Rankings}\label{sec:CompleteRankings}
We illustrate the case of complete rankings with the potato datasets described in \citet[Section 4]{Liu2019}. In short, a bag of 20 potatoes was bought, and 12 assessors were asked to rank the potatoes by weight, first by visual inspection, and next by holding the potatoes in hand. These datasets are available in \pkg{BayesMallows} as matrices with names \code{potato_weighing} and \code{potato_visual}, respectively. The true ranking of the potatoes' weights is available in the vector \code{potato_true_ranking}. 

The rankings given to potatoes 1-10 by assessors 1-4 in the visual experiment is printed below.
\begin{Schunk}
\begin{Sinput}
R> potato_visual[1:4, 1:10]
\end{Sinput}
\begin{Soutput}
   P1 P2 P3 P4 P5 P6 P7 P8 P9 P10
A1 10 18 19 15  6 16  4 20  3   5
A2 10 18 19 17 11 15  6 20  4   3
A3 12 15 18 16 13 11  7 20  6   3
A4  9 17 19 16 10 15  5 20  3   4
\end{Soutput}
\end{Schunk}
In general, \code{compute_mallows} expects ranking datasets to have one row for each assessor and one column for each item. Each row has to be a proper permutation, possibly with missing values. We are interested in both the level of agreement between assessors, as described by $\alpha$, and in the posterior latent ranking of the potatoes, as described by $\bm{\rho}$.

\subsubsection{Convergence Diagnostics}
\begin{figure}[t!]
\centering
\includegraphics[width=0.49\columnwidth]{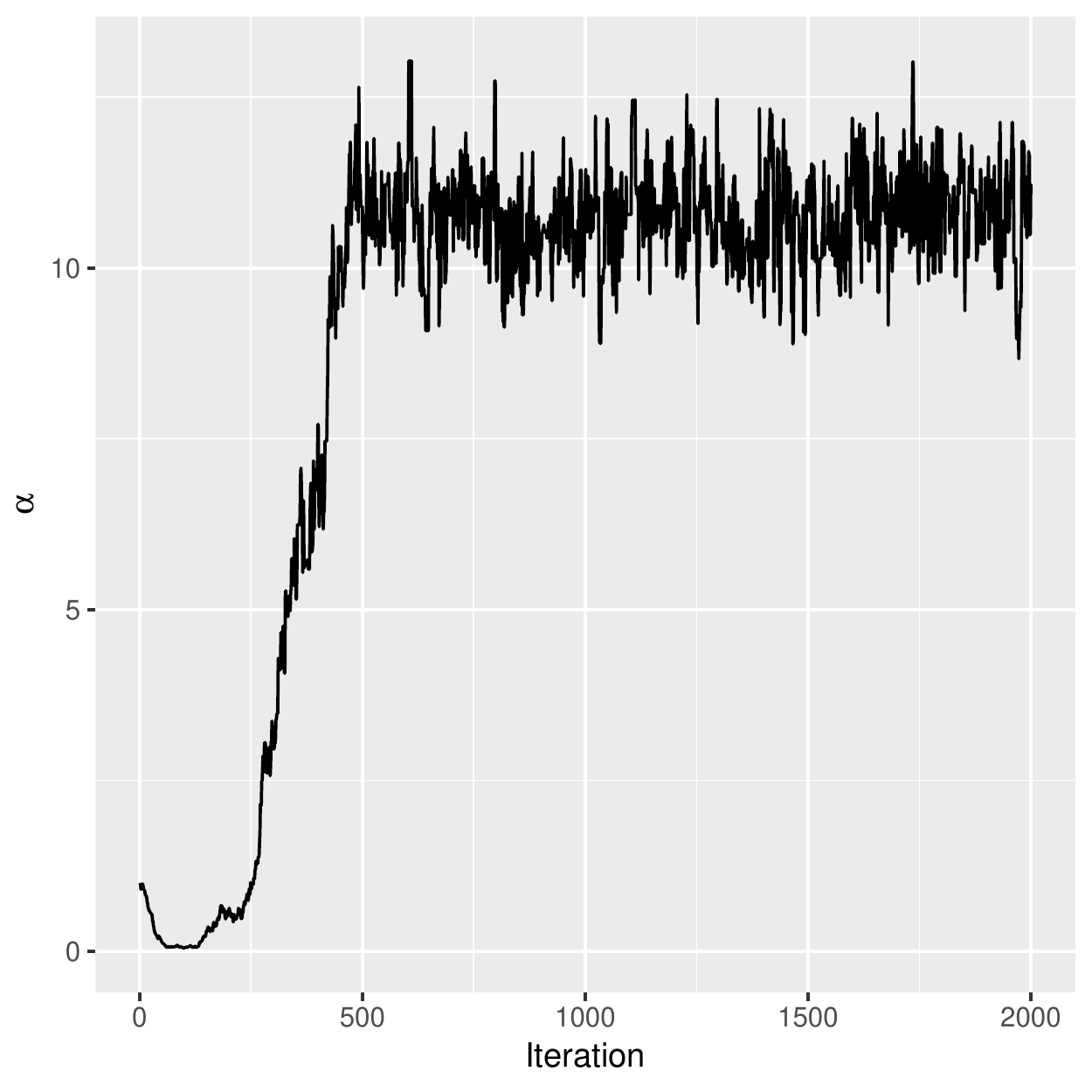}
\includegraphics[width=0.49\columnwidth]{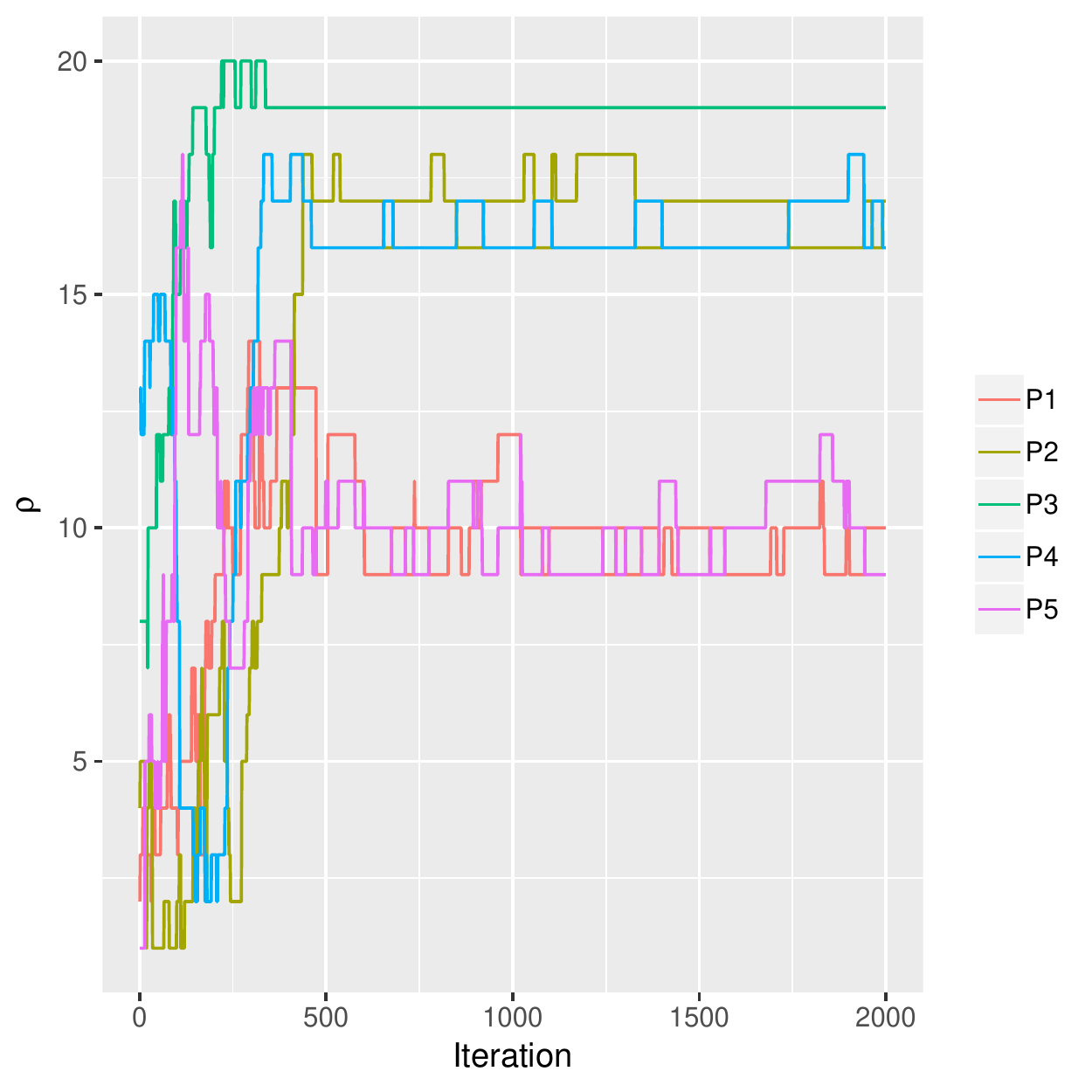}
\caption{\label{fig:potato_trace} Trace plot of $\alpha$ (left) and $\bm{\rho}$ (right) for the MCMC algorithm with the \code{potato\_visual} dataset.}
\end{figure}

First, we do a test run to check convergence of the MCMC algorithm. 
\begin{Schunk}
\begin{Sinput}
R> bmm_test <- compute_mallows(potato_visual)
\end{Sinput}
\end{Schunk}
Here, and in all subsequent code, we refer to the attached replication script for random number seeds for exact reproducibility. We get trace plots with \code{assess_convergence}.
\begin{Schunk}
\begin{Sinput}
R> assess_convergence(bmm_test)
\end{Sinput}
\end{Schunk}
By default, \code{assess_convergence} returns a trace plot for $\alpha$, shown in Figure~\ref{fig:potato_trace} (left). The algorithm seems to be mixing well after around 500 iterations. Next, we study the convergence of $\bm{\rho}$. To avoid too complicated plots, we pick potatoes $1-5$ by specifying this in the \code{items} argument.
\begin{Schunk}
\begin{Sinput}
R> assess_convergence(bmm_test, parameter = "rho", items = 1:5)
\end{Sinput}
\end{Schunk}
The corresponding plot is shown in Figure~\ref{fig:potato_trace} (right), illustrating that the MCMC algorithm seems to have converged after around 1000 iterations.

\subsubsection{Posterior Distributions}
\begin{figure}[t!]
\centering
\includegraphics[width=0.49\columnwidth]{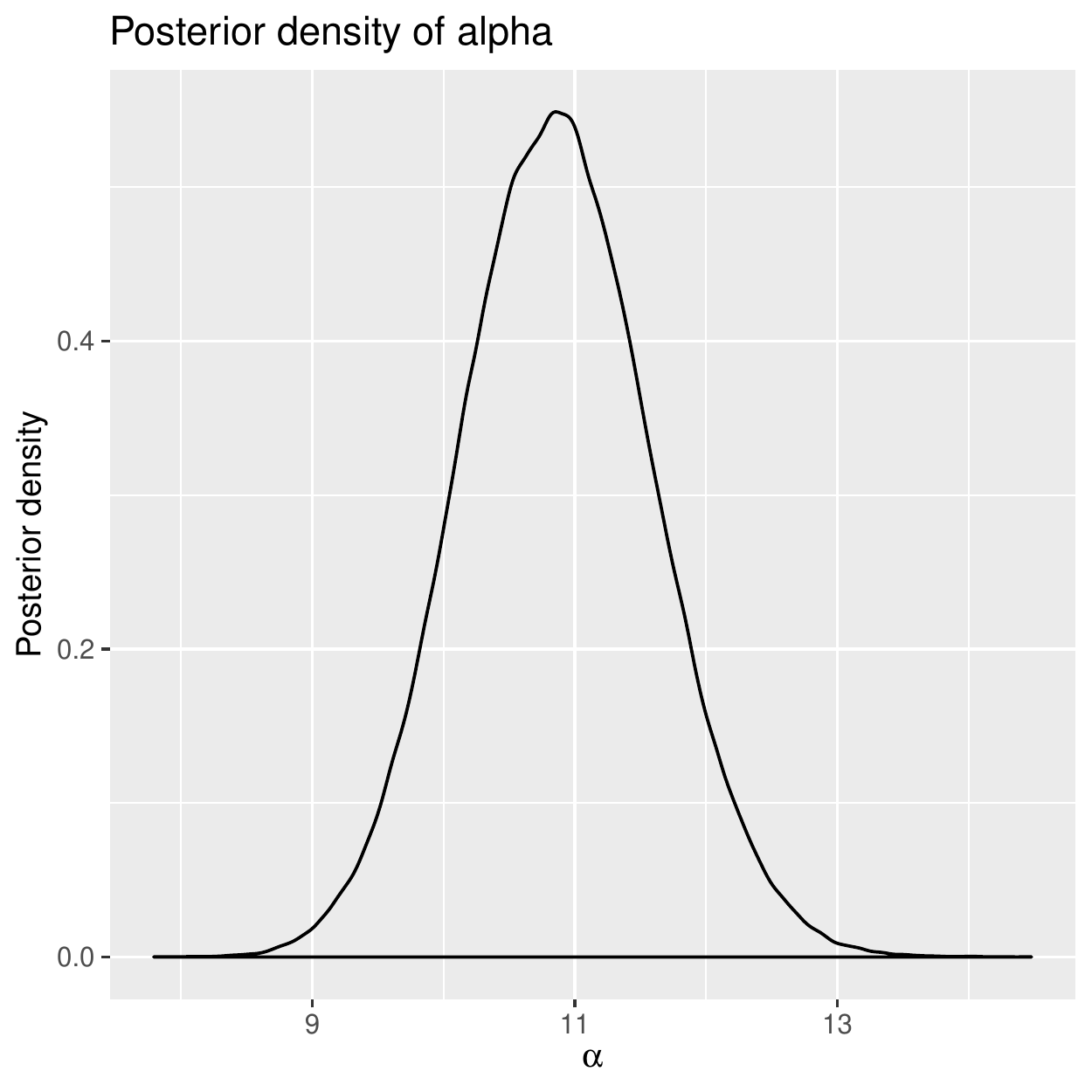}
\caption{\label{fig:potato_posterior_alpha} Posterior distribution of $\alpha$ with the \code{potato\_visual} dataset.}
\end{figure}

From the trace plots, we decide to discard the first 1000 MCMC samples as burn-in. We rerun the algorithm to get 500000 samples after burn-in.
\begin{Schunk}
\begin{Sinput}
R> bmm_visual <- compute_mallows(potato_visual, nmc = 501000)
\end{Sinput}
\end{Schunk}
After confirming that this new run shows similar convergence properties to the test run, we set the burn-in with the following command.
\begin{Schunk}
\begin{Sinput}
R> bmm_visual$burnin <- 1000
\end{Sinput}
\end{Schunk}
The object \code{bmm_visual} has \code{S3} class \code{BayesMallows}, so we plot the posterior distribution of $\alpha$ with \code{plot.BayesMallows} using
\begin{Schunk}
\begin{Sinput}
R> plot(bmm_visual)
\end{Sinput}
\end{Schunk}
The plot is shown in Figure~\ref{fig:potato_posterior_alpha}.

We can also get posterior credible intervals for $\alpha$ using \code{compute_posterior_intervals}, which returns both highest posterior density intervals (HPDI) and central intervals in a \code{tibble} \citep{Muller2018}.
\begin{Schunk}
\begin{Sinput}
R> compute_posterior_intervals(bmm_visual, decimals = 1L)
\end{Sinput}
\end{Schunk}
\begin{Schunk}
\begin{Soutput}
# A tibble: 1 x 6
  parameter  mean median conf_level hpdi       central_interval
  <chr>     <dbl>  <dbl> <chr>      <chr>      <chr>           
1 alpha      10.9   10.9 95 %       [9.4,12.3] [9.5,12.3]      
\end{Soutput}
\end{Schunk}
\pkg{BayesMallows} uses \code{tibble}s rather than \pkg{base} \proglang{R} \code{data.frame}s, but both are accepted as function inputs. We refer to \code{tibble}s as \emph{dataframes} henceforth. 

\begin{figure}[t!]
\centering
\includegraphics{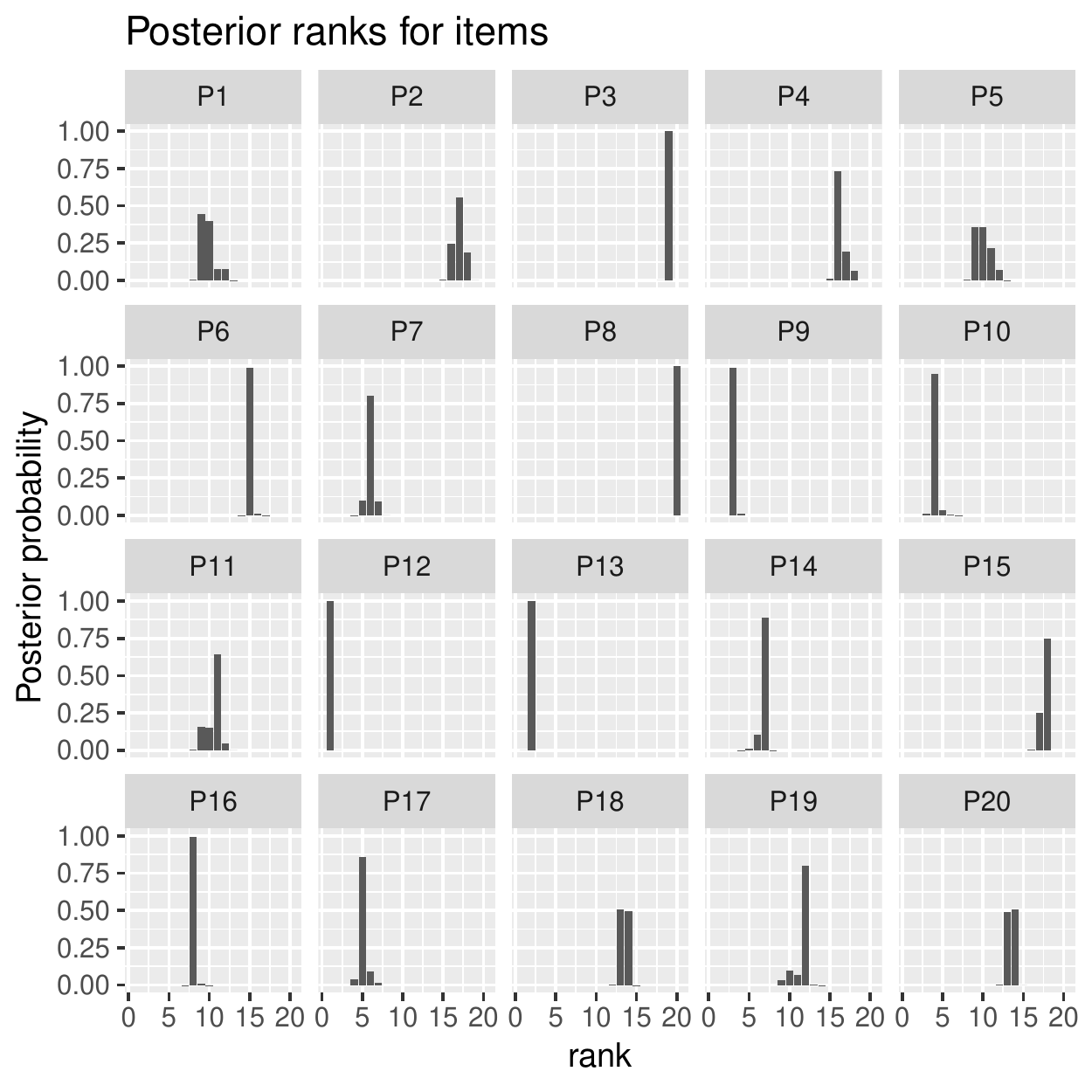}
\caption{\label{fig:potato_posterior_rho} Posterior distribution of latent ranks $\bm{\rho}$ with the \code{potato\_visual} dataset.}
\end{figure}

Next, we can go on to study the posterior distribution of $\bm{\rho}$. 
\begin{Schunk}
\begin{Sinput}
R> plot(bmm_visual, parameter = "rho", items = 1:20)
\end{Sinput}
\end{Schunk}
If the \code{items} argument is not provided, and the number of items exceeds five, five items are picked at random for plotting. To show all potatoes, we explicitly set \code{items = 1:20}. The corresponding plots are shown in Figure~\ref{fig:potato_posterior_rho}.

\subsubsection{Jumping Over the Scale Parameter}
Updating $\alpha$ in every step of the MCMC algorithm may not be necessary. With the \code{alpha_jump} argument, we can tell the MCMC algorithm to update $\alpha$ only every \code{alpha_jump}-th iteration. To update $\alpha$ every 10th update of $\bm{\rho}$ is updated, we do
\begin{Schunk}
\begin{Sinput}
R> bmm <- compute_mallows(potato_visual, nmc = 501000, alpha_jump = 10)
\end{Sinput}
\end{Schunk}
On a MacBook Pro 2.2 GHz Intel Core i7 running \proglang{R} version 3.5.1, the above call ran in 2.0 seconds on average over 1,000 replications using \pkg{microbenchmark} \citep{Mersmann2018}, while it took 4.2 seconds using the default value \code{alpha_jump = 1}, i.e., updating $\alpha$ less frequently more than halved the computing time.

\subsubsection{Other Distance Metrics}

By default, \code{compute_mallows} uses the footrule distance, but the user can also choose to use Cayley, Kendall, Hamming, Spearman, or Ulam distance. Running the same analysis of the potato data with Spearman distance is done with the command
\begin{Schunk}
\begin{Sinput}
R> bmm <- compute_mallows(potato_visual, 
+                         metric = "spearman", nmc = 501000)
\end{Sinput}
\end{Schunk}
For the particular case of Spearman distance, \pkg{BayesMallows} only has integer sequences for computing the exact partition function with 14 or fewer items. In this case a precomputed importance sampling estimated is part of the package, and used instead.

\subsection{Preference Data}\label{sec:PreferenceData}
Unless the argument \code{error_model} to \code{compute_mallows} is set, pairwise preference data are assumed to be consistent within each assessor. These data should be provided in a dataframe with the following three columns, with one row per pairwise comparison.
\begin{itemize}
\item \code{assessor} is an identifier for the assessor; either a numeric vector containing the assessor index, or a character vector containing the unique name of the assessor.
\item \code{bottom_item} is a numeric vector containing the index of the item that was disfavored in each pairwise comparison.
\item \code{top_item} is a numeric vector containing the index of the item that was preferred in each pairwise comparison.
\end{itemize}
A dataframe with this structure can be given in the \code{preferences} argument to \code{compute_mallows}. \code{compute_mallows} will generate the full set of implied rankings for each assessor using the function \code{generate_transitive_closure}, as well as an initial ranking matrix consistent with the pairwise preferences, using the function \code{generate_initial_ranking}.

We illustrate with the beach preference data containing stated pairwise preferences between random subsets of 15 images of beaches, by 60 assessors \citep[Section 6.2]{Vitelli2018}. This dataset is provided in the dataframe \code{beach_preferences}, and we can show the first 10 rows with the command
\begin{Schunk}
\begin{Sinput}
R> head(beach_preferences)
\end{Sinput}
\begin{Soutput}
# A tibble: 6 x 3
  assessor bottom_item top_item
     <dbl>       <dbl>    <dbl>
1        1           2       15
2        1           5        3
3        1          13        3
4        1           4        7
5        1           5       15
6        1          12        6
\end{Soutput}
\end{Schunk}

\subsubsection{Transitive Closure and Initial Ranking}
We start by generating the transitive closure implied by the pairwise preferences.
\begin{Schunk}
\begin{Sinput}
R> beach_tc <- generate_transitive_closure(beach_preferences)
\end{Sinput}
\end{Schunk}
The resulting object is a dataframe with subclass \code{BayesMallowsTC}, specifying that it is a transitive closure. \code{beach_tc} contains all the orderings in \code{beach_preferences}, but in addition all the implied orderings. Hence, the latter has 1442 rows, while the former has 2921 rows. We can now generate an initial ranking which is consistent with the implied orderings.
\begin{Schunk}
\begin{Sinput}
R> beach_init_rank <- generate_initial_ranking(beach_tc)
\end{Sinput}
\end{Schunk}
If we had not generated the transitive closure and the initial ranking, \code{compute_mallows} would do this for us, but when calling \code{compute_mallows} repeatedly, it may save time to have these precomputed and saved for future re-use. In order to save time in the case of big datasets, the functions for generating transitive closures and initial rankings from transitive closures can all be run in parallel, as shown in the examples to the \code{compute_mallows} function. The key to the parallelization is that each assessor's preferences can be handled independently of the others, and this can speed up the process considerably with large dataset.

As an example, let us look at all preferences stated by assessor 1 involving beach 2. We use \code{filter} from \pkg{dplyr} \citep{Wickham2018dplyr} to obtain the right set of rows.
\begin{Schunk}
\begin{Sinput}
R> library("dplyr")
R> filter(beach_preferences,
+         assessor == 1, bottom_item == 2 | top_item == 2)
\end{Sinput}
\begin{Soutput}
# A tibble: 1 x 3
  assessor bottom_item top_item
     <dbl>       <dbl>    <dbl>
1        1           2       15
\end{Soutput}
\end{Schunk}
Assessor 1 has performed only one direct comparison involving beach 2, in which the assessor stated that beach 15 is preferred to beach 2. The implied orderings, on the other hand, contain two preferences involving beach 2:
\begin{Schunk}
\begin{Sinput}
R> filter(beach_tc,
+         assessor == 1, bottom_item == 2 | top_item == 2)
\end{Sinput}
\begin{Soutput}
  assessor bottom_item top_item
1        1           2        6
2        1           2       15
\end{Soutput}
\end{Schunk}
In addition to the statement that beach 15 is preferred to beach 2, all the other orderings stated by assessor 1 imply that this assessor prefers beach 6 to beach 2.

\subsubsection{Convergence Diagnostics}
\begin{figure}[t!]
\centering
\includegraphics[width=0.49\columnwidth]{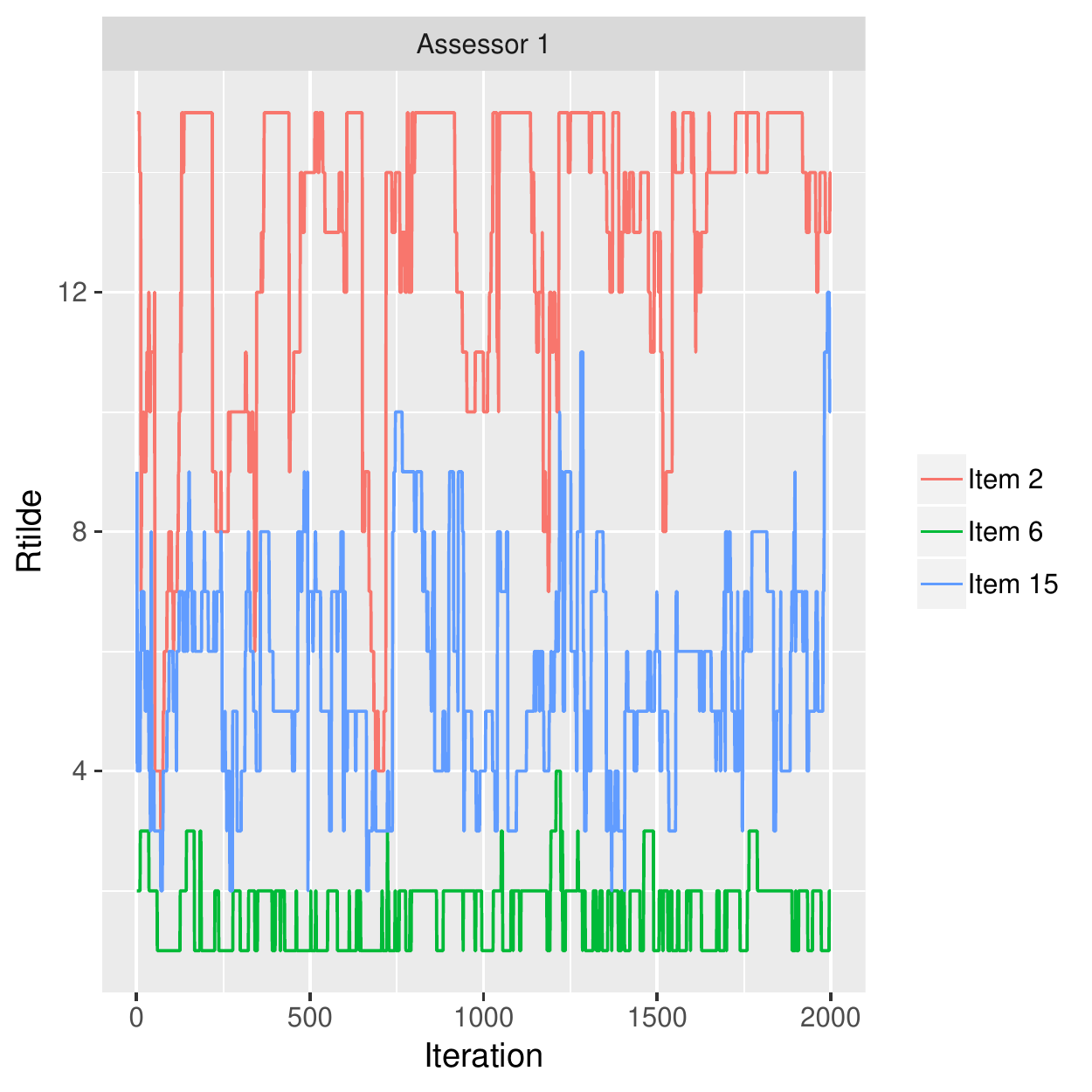}
\includegraphics[width=0.49\columnwidth]{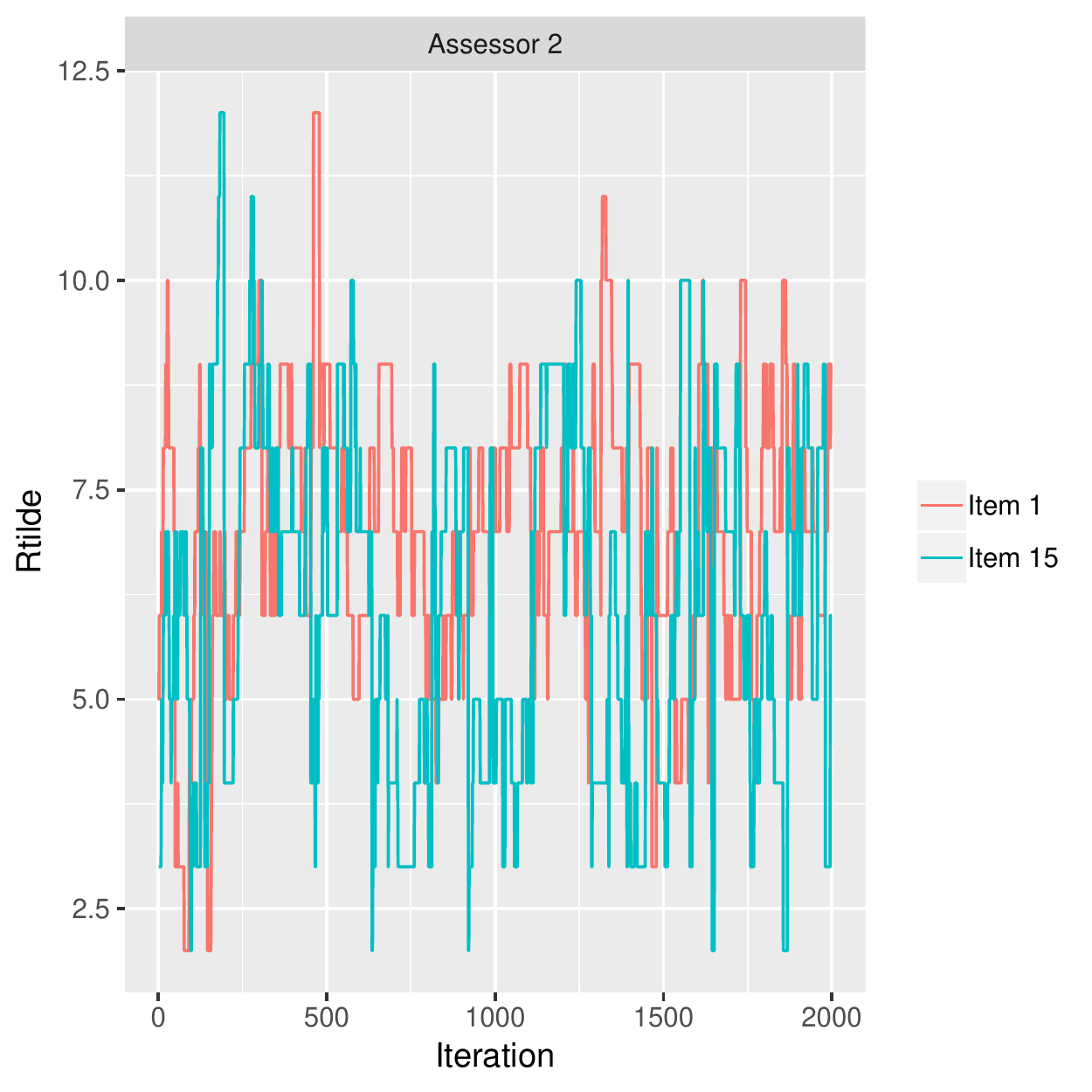}
\caption{\label{fig:beach_trace} Trace plots of augmented ranks $\tilde{R}$ for beaches 2, 6, and 15 for assessor 1 (left) and beaches 1 and 15 for assessor 2 (right).}
\end{figure}

As with the potato data, we can do a test run to assess the convergence of the MCMC algorithm. However, this time we provide the initial rankings \code{beach_init_rank} to the \code{rankings} argument and the transitive closure \code{beach_tc} to the \code{preferences} argument of \code{compute_mallows}. We also set \code{save_aug = TRUE} to save the augmented rankings in each MCMC step, hence letting us assess the convergence of the augmented rankings.
\begin{Schunk}
\begin{Sinput}
R> bmm_test <- compute_mallows(rankings = beach_init_rank,
+                              preferences = beach_tc, save_aug = TRUE)
\end{Sinput}
\end{Schunk}
Running \code{assess_convergence} for $\alpha$ and $\bm{\rho}$ shows good convergence after 1000 iterations (not shown). To check the convergence of the data augmentation scheme, we need to set \code{parameter = "Rtilde"}, and also specify which items and assessors to plot. Let us start by considering items 2, 6, and 15 for assessor 1, which we studied above.
\begin{Schunk}
\begin{Sinput}
R> assess_convergence(bmm_test, parameter = "Rtilde",
+                     items = c(2, 6, 15), assessors = 1)
\end{Sinput}
\end{Schunk}
The resulting plot is shown in Figure~\ref{fig:beach_trace} (left). It illustrates how the augmented rankings vary, while also obeying their implied ordering. 

By further investigation of \code{beach_tc}, we would find that no orderings are implied between beach 1 and beach 15 for assessor 2. With the following command, we create trace plots to confirm this:
\begin{Schunk}
\begin{Sinput}
R> assess_convergence(bmm_test, parameter = "Rtilde",
+                     items = c(1, 15), assessors = 2)
\end{Sinput}
\end{Schunk}
The resulting plot is shown in Figure~\ref{fig:beach_trace} (right). As expected, the traces of the augmented rankings for beach 1 and 15 for assessor 2 do cross each other, since no ordering is implied between them.

\begin{figure}[t!]
\centering
\includegraphics{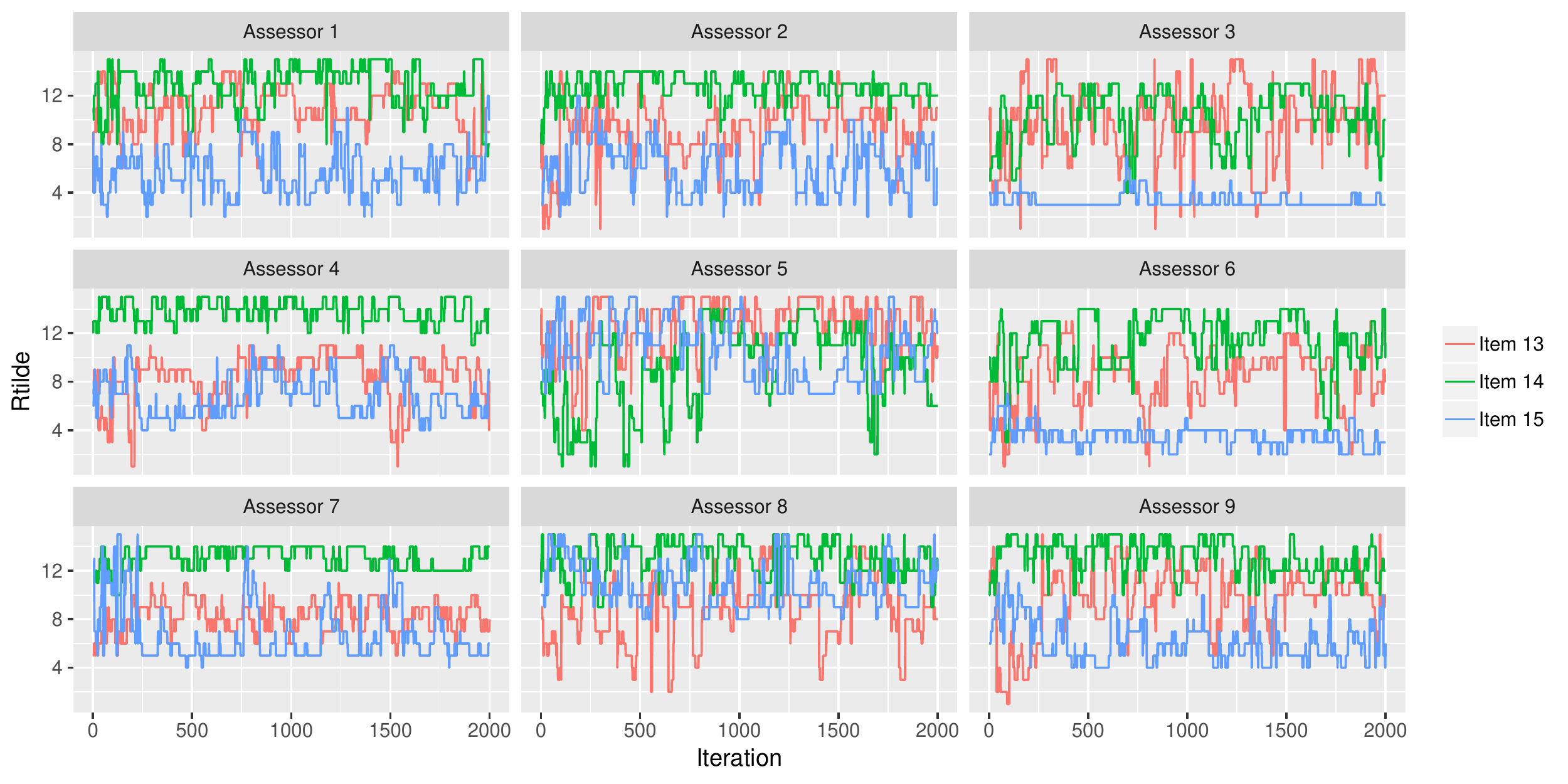}
\caption{\label{fig:beach_trace_all} Trace plots of augmented ranks $\tilde{R}$ for beaches 13-15 and assessors 1-9.}
\end{figure}

Ideally, we should look at trace plots for augmented ranks for more assessors to be sure that the algorithm is close to convergence. We can plot assessors 1-9 by setting \code{assessors = 1:9}. We also quite arbitrarily pick items 13-15, but the same procedure can be repeated for other items.
\begin{Schunk}
\begin{Sinput}
R> assess_convergence(bmm_test, parameter = "Rtilde",
+                     items = 13:15, assessors = 1:9)
\end{Sinput}
\end{Schunk}
The resulting plot is shown in Figure~\ref{fig:beach_trace_all}, indicating good mixing.

\subsubsection{Posterior Distributions}

Based on the convergence diagnostics, and being fairly conservative, we discard the first 2,000 MCMC iterations as burn-in, and take 100,000 additional samples.
\begin{Schunk}
\begin{Sinput}
R> bmm_beaches <- compute_mallows(
+    rankings = beach_init_rank,
+    preferences = beach_tc,
+    nmc = 102000,
+    save_aug = TRUE
+  )
R> bmm_beaches$burnin <- 2000
\end{Sinput}
\end{Schunk}
The posterior distributions of $\alpha$ and $\bm{\rho}$ can be studied as shown in Section~\ref{sec:CompleteRankings}. Posterior intervals for the latent rankings of each beach are obtained with
\begin{Schunk}
\begin{Sinput}
R> compute_posterior_intervals(bmm_beaches, parameter = "rho")
\end{Sinput}
\end{Schunk}
\begin{Schunk}
\begin{Soutput}
# A tibble: 15 x 7
   item    parameter  mean median conf_level hpdi    central_interval
   <fct>   <chr>     <dbl>  <dbl> <chr>      <chr>   <chr>           
 1 Item 1  rho           7      7 95 %       [7]     [6,7]           
 2 Item 2  rho          15     15 95 %       [15]    [15]            
 3 Item 3  rho           3      3 95 %       [3,4]   [3,4]           
 4 Item 4  rho          12     12 95 %       [11,13] [11,14]         
 5 Item 5  rho           9      9 95 %       [8,10]  [8,10]          
 6 Item 6  rho           2      2 95 %       [1,2]   [1,2]           
 7 Item 7  rho           9      8 95 %       [8,10]  [8,10]          
 8 Item 8  rho          12     11 95 %       [11,13] [11,14]         
 9 Item 9  rho           1      1 95 %       [1,2]   [1,2]           
10 Item 10 rho           6      6 95 %       [5,6]   [5,7]           
11 Item 11 rho           4      4 95 %       [3,4]   [3,5]           
12 Item 12 rho          13     13 95 %       [12,14] [12,14]         
13 Item 13 rho          10     10 95 %       [8,10]  [8,10]          
14 Item 14 rho          13     14 95 %       [12,14] [11,14]         
15 Item 15 rho           5      5 95 %       [5,6]   [4,6]           
\end{Soutput}
\end{Schunk}
We can also rank the beaches according to their cumulative probability (CP) consensus \citep[Section 5.1]{Vitelli2018} and their maximum posterior (MAP) rankings. This is done with the function \code{compute_consensus}, and the following call returns the CP consensus:
\begin{Schunk}
\begin{Sinput}
R> compute_consensus(bmm_beaches, type = "CP")
\end{Sinput}
\end{Schunk}
\begin{Schunk}
\begin{Soutput}
# A tibble: 15 x 3
   ranking item    cumprob
     <dbl> <chr>     <dbl>
 1       1 Item 9    0.896
 2       2 Item 6    1    
 3       3 Item 3    0.738
 4       4 Item 11   0.966
 5       5 Item 15   0.953
 6       6 Item 10   0.971
 7       7 Item 1    1    
 8       8 Item 7    0.528
 9       9 Item 5    0.887
10      10 Item 13   1.000
11      11 Item 8    0.508
12      12 Item 4    0.717
13      13 Item 12   0.643
14      14 Item 14   0.988
15      15 Item 2    1    
\end{Soutput}
\end{Schunk}
The column \code{cumprob} shows the probability of having the given rank or lower. Looking at the second row, for example, this means that beach 6 has probability 1 of having latent ranking 2 or lower. Next, beach 3 has probability 0.738 of having latent rank 3 or lower. This is an example of how the Bayesian framework can be used to not only rank items, but also to give posterior assessments of the uncertainty of the rankings. The MAP consensus is obtained similarly, by setting \code{type = "MAP"}.

\begin{figure}[t!]
\centering
\includegraphics{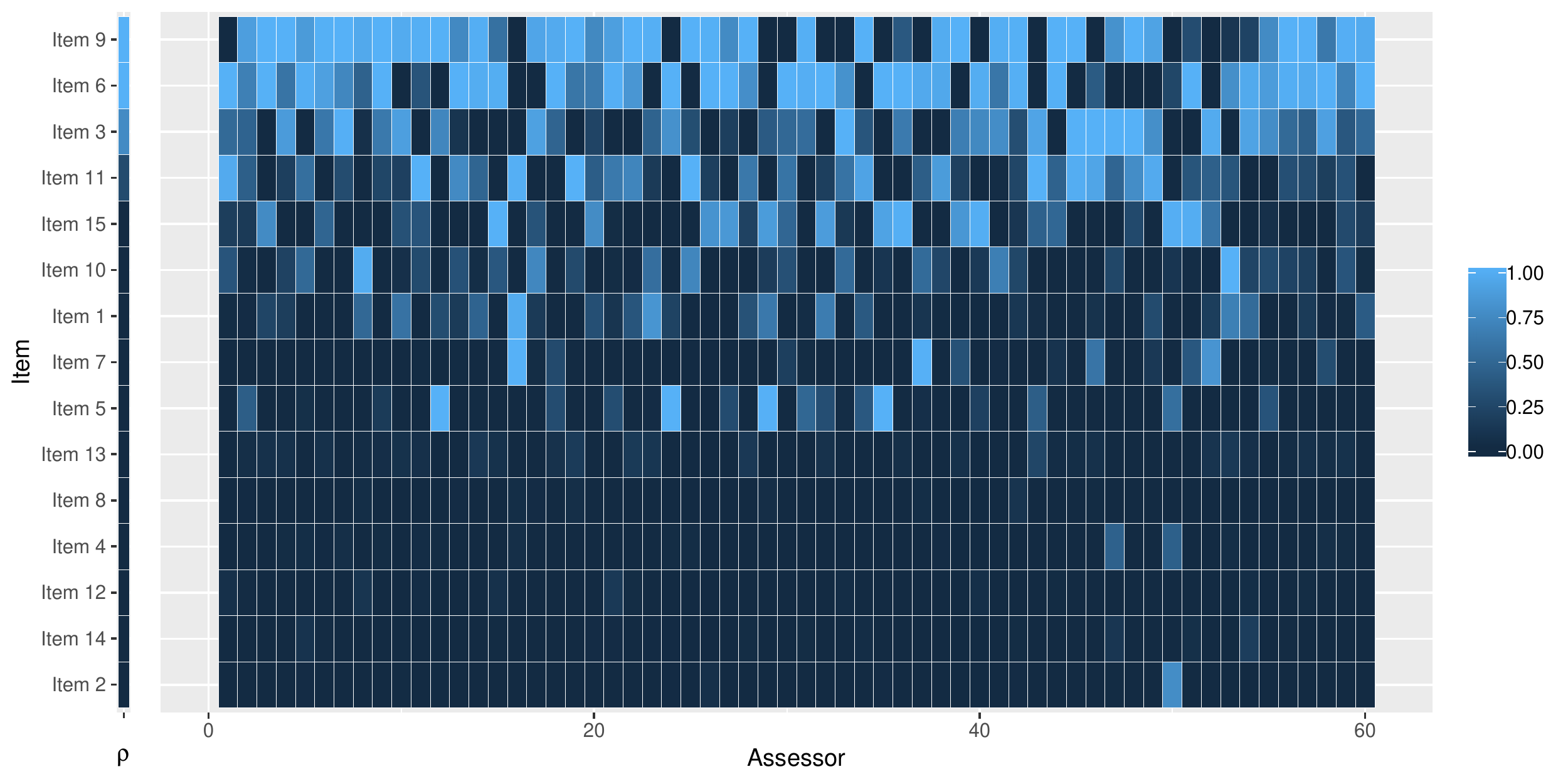}
\caption{\label{fig:beaches_top_3} Probability of being ranked top-3 for each beach in the beach preference example (left) and the probability that each assessor ranks the given beach among top-3 (right).}
\end{figure}

Keeping in mind that the ranking of beaches is based on sparse pairwise preferences, we can also ask: for beach $i$, what is the probability of being ranked top-$k$ by assessor $j$, and what is the probability of having latent rank among the top-$k$. The function \code{plot_top_k} plots these probabilities. By default, it sets \code{k = 3}, so a heatplot of the probability of being ranked top-3 is obtained with the call:
\begin{Schunk}
\begin{Sinput}
R> plot_top_k(bmm_beaches)
\end{Sinput}
\end{Schunk}
The plot is shown in Figure~\ref{fig:beaches_top_3}. The left part of the plot shows the beaches ranked according to their CP consensus, and the probability the latent rank $P(\rho_{i}) \leq 3$ for each beach $i$. The right part of the plot shows, for each beach as indicated on the left axis, the probability that assessor $j$ ranks the beach among top-3. For example, we see that assessor 1 has a very low probability of ranking beach 9 among her top-3, while assessor 3 has a very high probability of doing this.

The function \code{predict_top_k} returns a dataframe with all the underlying probabilities. For example, in order to find all the beaches that are among the top-3 of assessors 1-5 with more than 90 \% probability, we would do:
\begin{Schunk}
\begin{Sinput}
R> predict_top_k(bmm_beaches) %>%
+    filter(prob > 0.9, assessor %in% 1:5)
\end{Sinput}
\end{Schunk}
\begin{Schunk}
\begin{Soutput}
# A tibble: 6 x 3
# Groups:   assessor [4]
  assessor item     prob
     <dbl> <chr>   <dbl>
1        1 Item 11 0.955
2        1 Item 6  0.997
3        3 Item 6  0.997
4        3 Item 9  1    
5        4 Item 9  1.000
6        5 Item 6  0.979
\end{Soutput}
\end{Schunk}
Note that assessor 2 does not appear in this table, i.e., there are no beaches for which we are at least 90 \% certain that the beach is among assessor 2's top-3. In the code snippet above, we used the \pkg{magrittr} \citep{Bache2014} pipe operator \code{\%>\%}, which is exported from \pkg{dplyr}.

\subsection{Clustering}
\pkg{BayesMallows} comes with a set of sushi preference data, in which 5000 assessors each have ranked a set of 10 types of sushi \citep{Kamishima2003}. Here are the first six rows of the dataset:
\begin{Schunk}
\begin{Sinput}
R> head(sushi_rankings, 6)
\end{Sinput}
\begin{Soutput}
     shrimp sea eel tuna squid sea urchin salmon roe egg fatty tuna
[1,]      2       8   10     3          4          1   5          9
[2,]      1       8    6     4         10          9   3          5
[3,]      2       8    3     4          6          7  10          1
[4,]      4       7    5     6          1          2   8          3
[5,]      4      10    7     5          9          3   2          8
[6,]      4       6    2    10          7          5   1          9
     tuna roll cucumber roll
[1,]         7             6
[2,]         7             2
[3,]         5             9
[4,]         9            10
[5,]         1             6
[6,]         8             3
\end{Soutput}
\end{Schunk}
It is interesting to see if we can find subsets of assessors with similar preferences. The sushi dataset was analyzed with the BMM by \citet{Vitelli2018}, but the results in that paper differ somewhat from those obtained here, due to a bug in the function that was used to sample cluster probabilities from the Dirichlet distribution.

\subsubsection{Convergence Diagnostics}

\pkg{BayesMallows} has a convenient function \code{compute_mallows_mixtures} for computing multiple Mallows models with different numbers of mixture components. The function returns a list of models of class \code{BayesMallowsMixtures}, in which each list element contains a model with a given number of mixture components. Its arguments are \code{n_clusters}, which specifies the number of mixture components to compute, an optional parameter \code{cl} which can be set to the return value of the \code{makeCluster} function in the \pkg{parallel} package \citep{R}, and an ellipsis (\code{...}) for passing on arguments to \code{compute_mallows}. 

\begin{figure}[t!]
\centering
\includegraphics{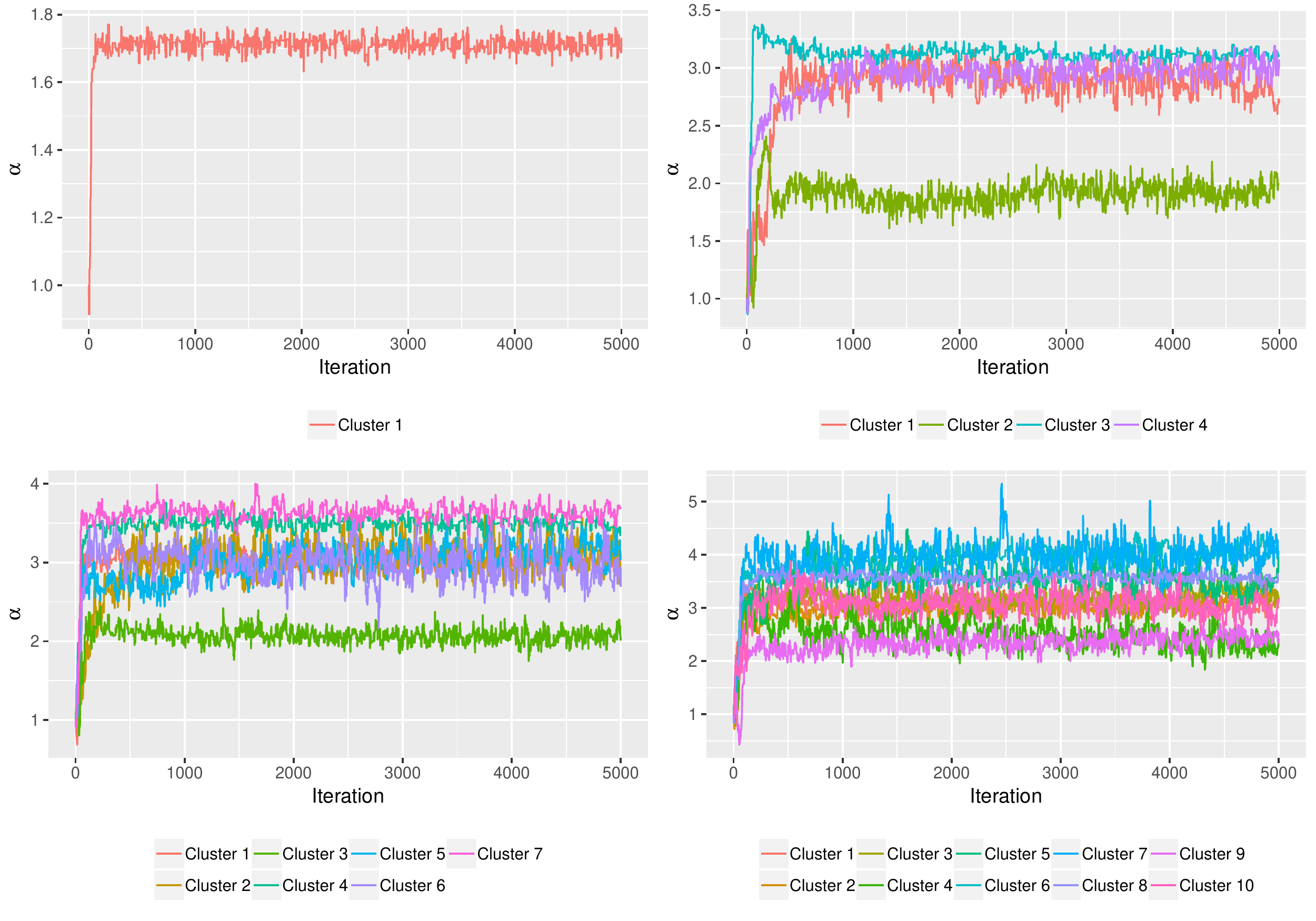}
\caption{\label{fig:sushi_alpha_trace} Trace plot of $\alpha$ for the sushi dataset with 1, 4, 7, and 10 mixture components, respectively.}
\end{figure}

Hypothesizing that we may not need more than 10 clusters to find a useful partitioning of the assessors, we start by doing test runs with 1, 4, 7, and 10 mixture components in order to assess convergence. We set the number of Monte Carlo samples to 5000, and since this is a test run, we do not save cluster assignments nor within-cluster distances from each MCMC iteration and hence set \code{save_clus = FALSE} and \code{include_wcd = FALSE}. We also run the computations in parallel on four cores, one for each mixture component.
\begin{Schunk}
\begin{Sinput}
R> library("parallel")
R> cl <- makeCluster(4)
R> bmm <- compute_mallows_mixtures(n_clusters = c(1, 4, 7, 10),
+                                  rankings = sushi_rankings, nmc = 5000,
+                                  save_clus = FALSE, include_wcd = FALSE,
+                                  cl = cl)
R> stopCluster(cl)
\end{Sinput}
\end{Schunk}
The function \code{assess_convergence} automatically creates a grid plot when given an object of class \code{BayesMallowsMixtures}, so we can check the convergence of $\alpha$ with the command
\begin{Schunk}
\begin{Sinput}
R> assess_convergence(bmm)
\end{Sinput}
\end{Schunk}
The resulting plot is given in Figure~\ref{fig:sushi_alpha_trace}, showing that all the chains seem to be close to convergence quite quickly.

\begin{figure}[t!]
\centering
\includegraphics{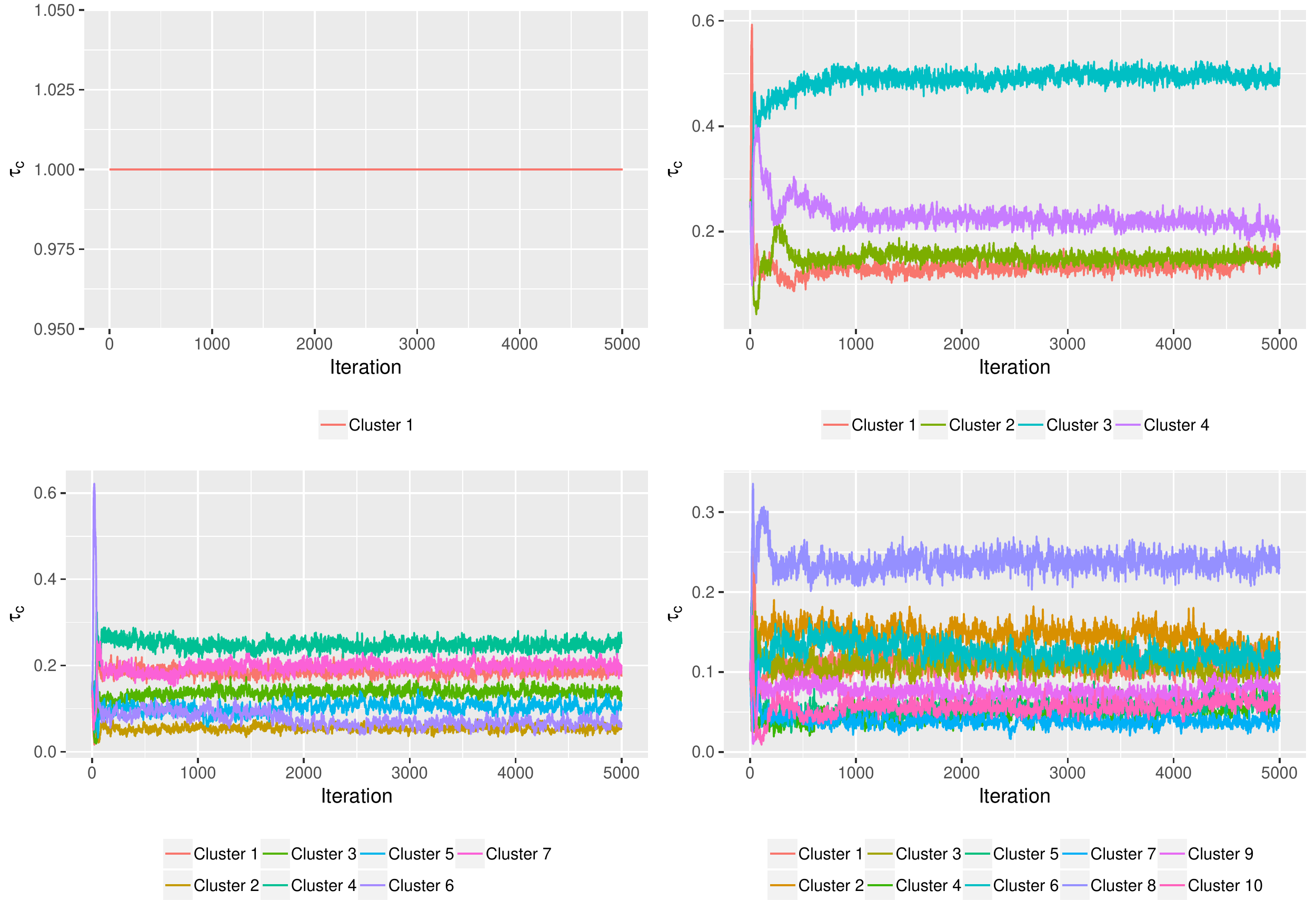}
\caption{\label{fig:sushi_tau_trace} Trace plot of $\tau_{c}$ for the sushi dataset with $c = 1$, $4$, $7$, and $10$ mixture components, respectively.}
\end{figure}

We can also make sure that the posterior distributions of the cluster probabilities $\tau_{c}$, $(c = 1, \dots, C)$ have converged properly, by setting \code{parameter = "cluster_probs"}. The trace plots for each number of mixture components are shown in Figure~\ref{fig:sushi_tau_trace}.
\begin{Schunk}
\begin{Sinput}
R> assess_convergence(bmm, parameter = "cluster_probs")
\end{Sinput}
\end{Schunk}
Note that with only one cluster, the cluster probability is fixed at the value 1 (upper left of Figure~\ref{fig:sushi_tau_trace}), while for other number of mixture components, the chains seem to be mixing well.

\subsubsection{Deciding on the Number of Mixtures}

\begin{figure}[t!]
\centering
\includegraphics{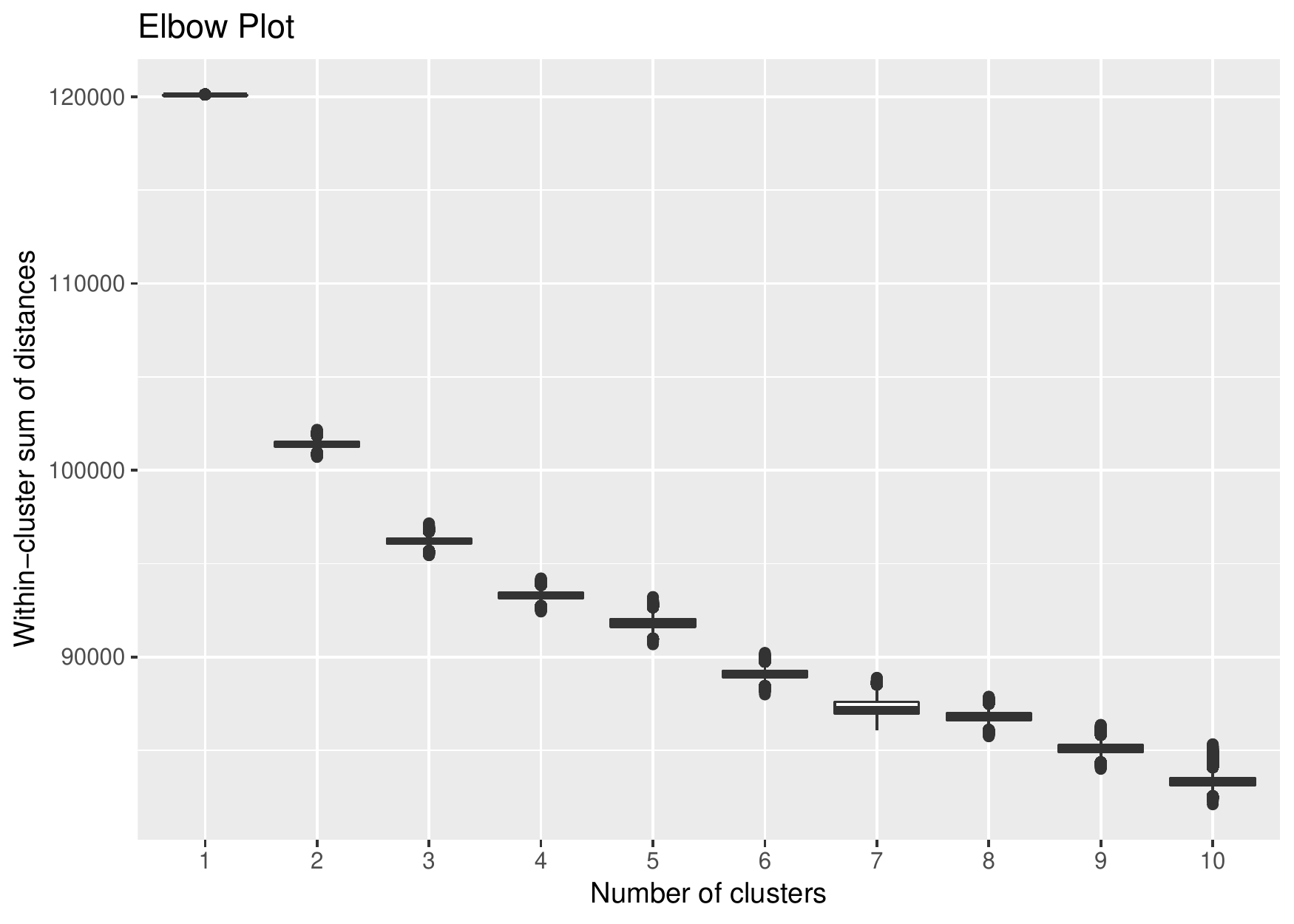}
\caption{\label{fig:sushi_elbow} Elbow plot for the sushi mixture models.}
\end{figure}

Given the convergence assessment of the previous section, we are fairly confident that a burn-in of 5,000 is sufficient. We run 95,000 additional iterations, and try from 1 to 10 mixture components. Our goal is now to determine the number of mixture components to use, and in order to create an elbow plot, we set \code{include_wcd = TRUE} to compute the within-cluster distances in each step of the MCMC algorithm. Since the posterior distributions of $\rho_{c}$ ($c = 1,\dots,C$) are highly peaked, we save some memory by only saving every 10th value of $\bm{\rho}$ by setting \code{rho_thinning = 10}.
\begin{Schunk}
\begin{Sinput}
R> cl <- makeCluster(4)
R> bmm <- compute_mallows_mixtures(n_clusters = 1:10, 
+                                  rankings = sushi_rankings, nmc = 100000, 
+                                  rho_thinning = 10, save_clus = FALSE, 
+                                  include_wcd = TRUE, cl = cl)
R> stopCluster(cl)
\end{Sinput}
\end{Schunk}
We create an elbow plot with the command:
\begin{Schunk}
\begin{Sinput}
R> plot_elbow(bmm, burnin = 5000)
\end{Sinput}
\end{Schunk}
The resulting plot is shown in Figure~\ref{fig:sushi_elbow}. Although not clear-cut, we see that the within-cluster sum of distances levels off at around 5 clusters, and hence choose to use 5 clusters in our model.

\subsubsection{Posterior Distributions}

\begin{figure}[t!]
\centering
\includegraphics{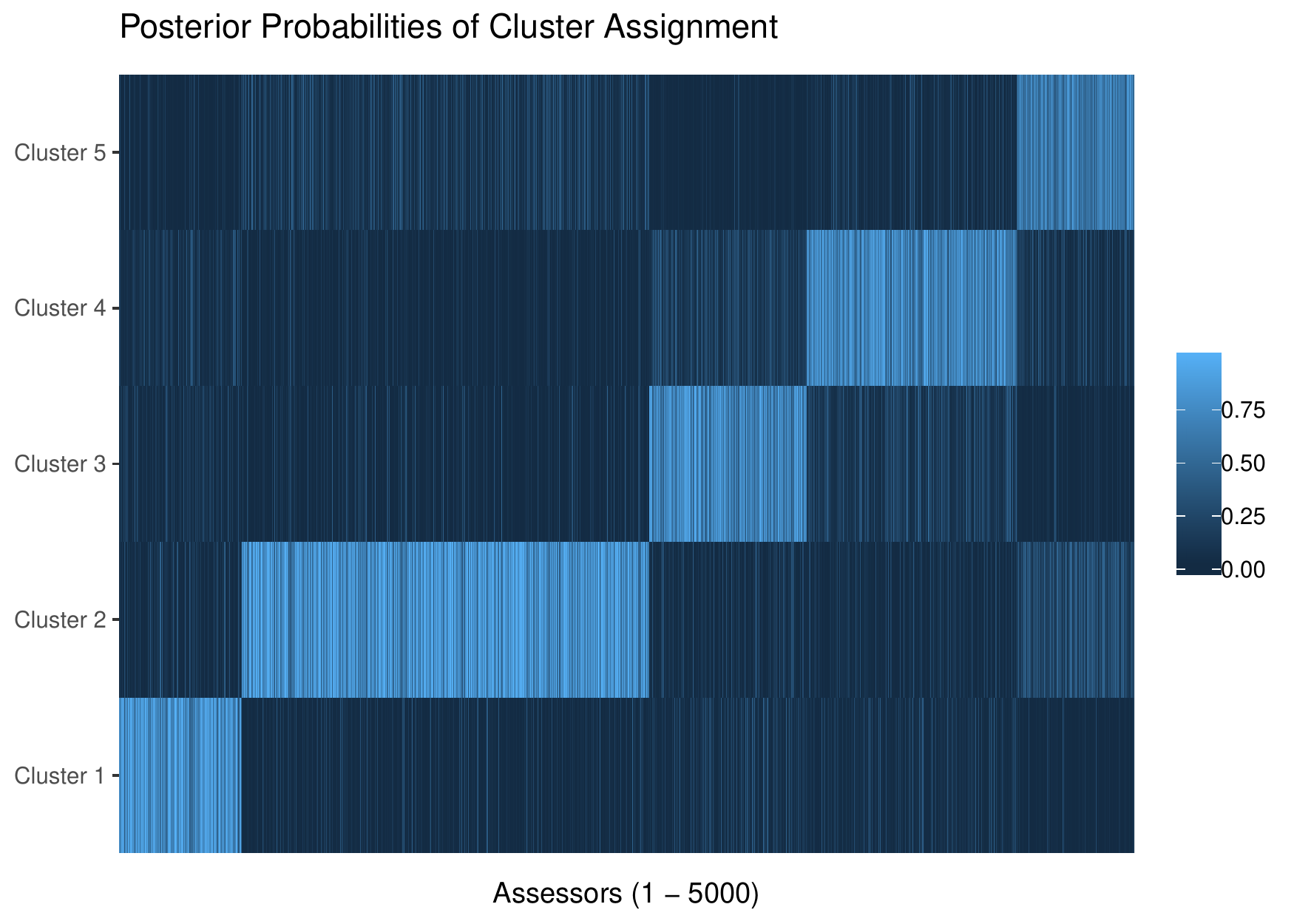}
\caption{\label{fig:sushi_cluster_assignment} Posterior probabilities of assignment to each cluster for each of the 5000 assessors in the sushi dataset. The scale to the right shows the color coding of probabilities.}
\end{figure}

Having chosen 5 mixture components, we go on to fit a final model, still running 95,000 iterations after burnin. This time we call \code{compute_mallows} and set \code{n_clusters = 5}. We also set \code{save_clus = TRUE} and \code{clus_thin = 10} to save the cluster assignments of each assessor in every 10th iteration, and \code{rho_thinning = 10} to save the estimated latent rank every 10th iteration.
\begin{Schunk}
\begin{Sinput}
R> bmm <- compute_mallows(rankings = sushi_rankings, n_clusters = 5, 
+                         save_clus = TRUE, clus_thin = 10, nmc = 100000,
+                         rho_thinning = 10)
\end{Sinput}
\end{Schunk}
The burnin is set to 5,000, as before.
\begin{Schunk}
\begin{Sinput}
R> bmm$burnin <- 5000
\end{Sinput}
\end{Schunk}
We can plot the posterior distributions of $\alpha$ and $\bm{\rho}$ in each cluster using \code{plot.BayesMallows} as shown above for the potato data in Section \ref{sec:CompleteRankings}. We can also show the posterior distributions of the cluster probabilities, using:
\begin{Schunk}
\begin{Sinput}
R> plot(bmm, parameter = "cluster_probs")
\end{Sinput}
\end{Schunk}
Using the argument \code{parameter = "cluster_assignment"}, we can visualize the posterior probability for each assessor of belonging to each cluster:
\begin{Schunk}
\begin{Sinput}
R> plot(bmm, parameter = "cluster_assignment")
\end{Sinput}
\end{Schunk}
The resulting plot is shown in Figure~\ref{fig:sushi_cluster_assignment}. The underlying numbers can be obtained using the function \code{assign_cluster}.

We can find clusterwise consensus rankings using \code{compute_consensus}. The following call finds the CP consensuses, and then uses \code{select} from \pkg{dplyr} \citep{Wickham2018dplyr} and \code{spread} from \pkg{tidyr} \citep{Wickham2018tidyr} to create one column for each cluster. The result is shown in Table~\ref{tab:sushi_consensus}.
\begin{Schunk}
\begin{Sinput}
R> library("tidyr")
R> compute_consensus(bmm) %>%
+    select(-cumprob) %>%
+    spread(key = cluster, value = item)
\end{Sinput}
\end{Schunk}
%
\begin{table}[ht]
\centering
\begin{tabular}{rlllll}
  \hline
 & Cluster 1 & Cluster 2 & Cluster 3 & Cluster 4 & Cluster 5 \\ 
  \hline
1 & shrimp & fatty tuna & fatty tuna & fatty tuna & fatty tuna \\ 
  2 & sea eel & sea urchin & sea eel & tuna & sea urchin \\ 
  3 & egg & salmon roe & tuna & shrimp & shrimp \\ 
  4 & squid & sea eel & shrimp & tuna roll & tuna \\ 
  5 & salmon roe & tuna & tuna roll & squid & salmon roe \\ 
  6 & fatty tuna & shrimp & squid & salmon roe & squid \\ 
  7 & tuna & tuna roll & egg & egg & tuna roll \\ 
  8 & tuna roll & squid & cucumber roll & cucumber roll & sea eel \\ 
  9 & cucumber roll & egg & salmon roe & sea eel & egg \\ 
  10 & sea urchin & cucumber roll & sea urchin & sea urchin & cucumber roll \\ 
   \hline
\end{tabular}
\caption{CP consensus for each of the clusters found for sushi data.} 
\label{tab:sushi_consensus}
\end{table}

Note that for estimating cluster specific parameters, label switching is a potential problem that needs to be handled. \pkg{BayesMallows} ignores label switching issues inside the MCMC, because it has been shown that this approach is better for ensuring full convergence of the chain \citep{Jasra2005, Celeux2000}. MCMC iterations can be re-ordered after convergence is achieved, for example by using the implementation of Stephens' algorithm \citep{Stephens2000} provided by the \proglang{R} package \pkg{label.switching} \citep{Papastamoulis2015}. A full example of how to assess label switching after running \code{compute_mallows} is provided by running the following command:
\begin{Schunk}
\begin{Sinput}
R> help("label_switching")
\end{Sinput}
\end{Schunk}
For the sushi data analyzed in this section, no label switching is detected by Stephen's algorithm.

\section{Discussion}\label{sec:concl}
In this paper we discussed the methodological background and  computational strategies for the \pkg{BayesMallows} package, implementing the inferential framework for the analysis of preference data based on the Bayesian Mallows model, as introduced in \citet{Vitelli2018}. The package aims at providing a general probabilistic tool, capable of performing various inferential tasks (estimation, classification, prediction) with a proper uncertainty quantification. Moreover, the package widens the applicability of the Mallows model, by providing reliable algorithms for approximating the associated partition function, which has been the bottleneck for a successful use of this general and flexible model so far. Finally, it handles a variety of preference data types (partial rankings, pairwise preferences), and it could possibly handle many others which can lie in the above mentioned categories (noisy continuous measurements, clicking data, ratings).

One of the most important features of the \pkg{BayesMallows} package is that, despite implementing a Bayesian model, and thus relying on MCMC algorithms, its clever implementation makes it actually possible to use it for large datasets. The package can easily handle up to hundreds of items, and thousands of assessors; an example is the Movielens data analyzed in Section 6.4 of \citep{Vitelli2018}. By using the log-sum-exp trick, the implementation of the importance sampler is able to handle at least ten thousand items without numerical overflow. We believe that all these features make the package a unique resource for fitting the Mallows model to large data, with the benefits of a fully probabilistic interpretation.

Nonetheless, we also recognize that the \pkg{BayesMallows} package can open the way for further generalizations. The Bayesian Mallows model for time-varying rankings that has been introduced in \citet{asfaw2017time} will be considered for a future release. Some further extensions which we might consider to implement in the \pkg{BayesMallows} in the future include: fitting an infinite mixture of Mallows models for automatically performing model selection; allowing for a non-uniform prior for $\bm{\rho}$; performing automatic item selection; estimating the assessors' quality as rankers; and finally including covariates, both on the assessors and on the items. In addition, since the data augmentation steps in the MCMC algorithm are independent across assessors, potential speedup in the case of missing data or pairwise preferences can be obtained by updating the augmented data in parallel, and this is likely to be part of a future package update.


\section*{Computational Details}
The M-H algorithm for computing the BMM, as well as the importance sampling algorithm for estimating the partition function, are implemented in \proglang{C++}, heavily utilizing the \code{Armadillo} library for linear algebra and scientific computing \citep{Sanderson2016,Sanderson2018}, using \pkg{Rcpp} \citep{Eddelbuettel2011,Eddelbuettel2013,Eddelbuettel2017} and \pkg{RcppArmadillo} \citep{Eddelbuettel2014} for integrating the \proglang{C++} code in \proglang{R}. For computing the Ulam distance, \pkg{BayesMallows} uses the function \code{perm0_dist} in the \proglang{C++} library \pkg{SUBSET} \citep{Burkardt2018}, and for computing the partition function for the MM with Ulam distance, the function \code{count.perms} from the \pkg{PerMallows} package \citep{Irurozki2016} and the algorithm outlined in \citet{Irurozki2014Ulam} is used.

The packages \pkg{relations} \citep{Meyer2018} and \pkg{sets} \citep{Meyer2009} are used for generating the transitive closure implied by pairwise preferences, while \pkg{igraph} is used for generating initial rankings based on such transitive closures.

\pkg{BayesMallows} heavily uses the packages \pkg{dplyr} \citep{Wickham2018dplyr}, \pkg{purrr} \citep{Henry2018purrr}, and \pkg{tidyr} \citep{Wickham2018tidyr} for data transformation both before and after running the MCMC algorithm, and \pkg{rlang} \citep{Henry2018rlang} is used for programming with these packages. All plots are produced with \pkg{ggplot2} \citep{Wickham2016}, sometimes with the help of \pkg{cowplot} \citep{Wilke2018} for aligning several plots in a grid. The Bayesian posterior intervals are computed using \pkg{HDInterval} \citep{Meredith2018}.

The \pkg{devtools} package provided very useful tools during package development, and the function \code{check_with_sanitizers} function in the \pkg{rhub} package \citep{Csardi2017} was useful for detecting memory issues in the \proglang{C++} code using address sanitizers and leak sanitizers as described in Sections 4.3.3 and 4.3.4 of Writing \proglang{R} Extensions \citet{RExt}.

\section*{Acknowledgments}

The authors would like to thank Arnoldo Frigessi and Elja Arjas for fruitful discussions.


\bibliography{references}

\begin{thebibliography}{48}
\newcommand{\enquote}[1]{``#1''}
\providecommand{\natexlab}[1]{#1}
\providecommand{\url}[1]{\texttt{#1}}
\providecommand{\urlprefix}{URL }
\expandafter\ifx\csname urlstyle\endcsname\relax
  \providecommand{\doi}[1]{doi:\discretionary{}{}{}#1}\else
  \providecommand{\doi}{doi:\discretionary{}{}{}\begingroup
  \urlstyle{rm}\Url}\fi
\providecommand{\eprint}[2][]{\url{#2}}

\bibitem[{Alvo and Yu(2014)}]{Alvo2014}
Alvo M, Yu PL (2014).
\newblock \emph{Statistical Methods for Ranking Data}.
\newblock Frontiers in Probability and the Statistical Sciences. Springer, New
  York, NY, USA.
\newblock \doi{10.1007/978-1-4939-1471-5}.

\bibitem[{Asfaw \emph{et~al.}(2017)Asfaw, Vitelli, S{\o}rensen, Arjas, and
  Frigessi}]{asfaw2017time}
Asfaw D, Vitelli V, S{\o}rensen {\O}, Arjas E, Frigessi A (2017).
\newblock \enquote{Time-Varying Rankings With the Bayesian Mallows Model.}
\newblock \emph{Stat}, \textbf{6}(1), 14--30.

\bibitem[{Bache and Wickham(2014)}]{Bache2014}
Bache SM, Wickham H (2014).
\newblock \emph{\pkg{magrittr}: A Forward-Pipe Operator for R}.
\newblock \proglang{R} package version 1.5,
  \urlprefix\url{https://CRAN.R-project.org/package=magrittr}.

\bibitem[{Burkardt(2018)}]{Burkardt2018}
Burkardt J (2018).
\newblock \emph{\pkg{SUBSET} Combinatorial Routines}.
\newblock
  \urlprefix\url{http://people.sc.fsu.edu/~jburkardt/cpp_src/subset/subset.html}.

\bibitem[{Celeux \emph{et~al.}(2000)Celeux, Hurn, and Robert}]{Celeux2000}
Celeux G, Hurn M, Robert C (2000).
\newblock \enquote{Computational and Inferential Difficulties with Mixture
  Posterior Distribution.}
\newblock \emph{Journal of the American Statistical Association},
  \textbf{95}(451), 957--970.

\bibitem[{Crispino \emph{et~al.}(2018)Crispino, Arjas, Vitelli, Barrett, and
  Frigessi}]{crispino2017bayesian}
Crispino M, Arjas E, Vitelli V, Barrett N, Frigessi A (2018).
\newblock \enquote{A Bayesian Mallows Approach to Non-Transitive Pair
  Comparison Data: How Human are Sounds?}
\newblock \emph{Forthcoming in the Annals of Applied Statistics}.

\bibitem[{Csárdi(2017)}]{Csardi2017}
Csárdi G (2017).
\newblock \emph{rhub: Connect to 'R-hub'}.
\newblock \proglang{R} package version 1.0.2,
  \urlprefix\url{https://CRAN.R-project.org/package=rhub}.

\bibitem[{de~Borda(1781)}]{Borda1781}
de~Borda JC (1781).
\newblock \enquote{M{\'e}moire sur les {\'e}lections au scrutin, Histoire de
  l'Acad{\'e}mie Royale des Sciences.}
\newblock \emph{Paris, France}.

\bibitem[{Diaconis(1988)}]{Diaconis1988}
Diaconis P (1988).
\newblock \emph{Group Representations in Probability and Statistics}, volume~11
  of \emph{Lecture Notes - Monograph Series}.
\newblock Institute of Mathematical Statistics, Hayward, CA, USA.

\bibitem[{Eddelbuettel(2013)}]{Eddelbuettel2013}
Eddelbuettel D (2013).
\newblock \emph{Seamless \proglang{R} and \proglang{C++} Integration with
  \pkg{Rcpp}}.
\newblock Springer, New York.
\newblock \doi{10.1007/978-1-4614-6868-4}.
\newblock ISBN 978-1-4614-6867-7.

\bibitem[{Eddelbuettel and Balamuta(2017)}]{Eddelbuettel2017}
Eddelbuettel D, Balamuta JJ (2017).
\newblock \enquote{Extending \proglang{R} with \proglang{C++}: A Brief
  Introduction to \pkg{Rcpp}.}
\newblock \emph{PeerJ Preprints}, \textbf{5}, e3188v1.
\newblock ISSN 2167-9843.
\newblock \doi{10.7287/peerj.preprints.3188v1}.
\newblock \urlprefix\url{https://doi.org/10.7287/peerj.preprints.3188v1}.

\bibitem[{Eddelbuettel and Francois(2011)}]{Eddelbuettel2011}
Eddelbuettel D, Francois R (2011).
\newblock \enquote{\pkg{Rcpp}: Seamless \proglang{R} and \proglang{C++}
  Integration.}
\newblock \emph{Journal of Statistical Software}, \textbf{40}, 1--18.
\newblock \urlprefix\url{http://www.jstatsoft.org/v40/i08/}.

\bibitem[{Eddelbuettel and Sanderson(2014)}]{Eddelbuettel2014}
Eddelbuettel D, Sanderson C (2014).
\newblock \enquote{\pkg{RcppArmadillo}: Accelerating \proglang{R} with
  High-Performance \proglang{C++} Linear Algebra.}
\newblock \emph{Computational Statistics and Data Analysis}, \textbf{71},
  1054--1063.
\newblock \urlprefix\url{http://dx.doi.org/10.1016/j.csda.2013.02.005}.

\bibitem[{Fligner and Verducci(1986)}]{Fligner1986}
Fligner MA, Verducci JS (1986).
\newblock \enquote{Distance Based Ranking Models.}
\newblock \emph{Journal of the Royal Statistical Society B}, \textbf{48}(3),
  359--369.
\newblock ISSN 00359246.
\newblock \urlprefix\url{http://www.jstor.org/stable/2345433}.

\bibitem[{Henry and Wickham(2018{\natexlab{a}})}]{Henry2018purrr}
Henry L, Wickham H (2018{\natexlab{a}}).
\newblock \emph{\pkg{purrr}: Functional Programming Tools}.
\newblock \proglang{R} package version 0.2.5,
  \urlprefix\url{https://CRAN.R-project.org/package=purrr}.

\bibitem[{Henry and Wickham(2018{\natexlab{b}})}]{Henry2018rlang}
Henry L, Wickham H (2018{\natexlab{b}}).
\newblock \emph{\pkg{rlang}: Functions for Base Types and Core \proglang{R} and
  'Tidyverse' Features}.
\newblock \proglang{R} package version 0.2.2,
  \urlprefix\url{https://CRAN.R-project.org/package=rlang}.

\bibitem[{Irurozki \emph{et~al.}(2014{\natexlab{a}})Irurozki, Calvo, and
  Lozano}]{irurozki2014Ham}
Irurozki E, Calvo B, Lozano A (2014{\natexlab{a}}).
\newblock \enquote{Sampling and Learning the Mallows and Weighted Mallows
  Models Under the Hamming Distance.}
\newblock \emph{Technical Report}.
\newblock \urlprefix\url{https://addi.ehu.es/handle/10810/11240}.

\bibitem[{Irurozki \emph{et~al.}(2016)Irurozki, Calvo, and
  Lozano}]{Irurozki2016}
Irurozki E, Calvo B, Lozano JA (2016).
\newblock \enquote{\pkg{PerMallows}: An \proglang{R} Package for Mallows and
  Generalized Mallows Models.}
\newblock \emph{Journal of Statistical Software}, \textbf{71}(12), 1--30.
\newblock \doi{10.18637/jss.v071.i12}.

\bibitem[{Irurozki \emph{et~al.}(2018)Irurozki, Calvo, and
  Lozano}]{irurozki2016sampling}
Irurozki E, Calvo B, Lozano JA (2018).
\newblock \enquote{Sampling and Learning the {M}allows and Generalized
  {M}allows Models Under the {C}ayley Distance.}
\newblock \emph{Methodology and Computing in Applied Probability},
  \textbf{20}(1), 1--35.

\bibitem[{Irurozki \emph{et~al.}(2014{\natexlab{b}})Irurozki, Ceberio, Calvo,
  and Lozano}]{Irurozki2014Ulam}
Irurozki E, Ceberio J, Calvo B, Lozano JA (2014{\natexlab{b}}).
\newblock \enquote{Sampling and Learning the Mallows Model Under the Ulam
  Distance.}
\newblock \emph{Technical Report}.
\newblock \urlprefix\url{https://addi.ehu.es/handle/10810/11241}.

\bibitem[{Jasra \emph{et~al.}(2005)Jasra, Holmes, and Stephens}]{Jasra2005}
Jasra A, Holmes C, Stephens D (2005).
\newblock \enquote{Markov Chain {M}onte {C}arlo Methods and the Label Switching
  Problem in {B}ayesian Mixture Modeling.}
\newblock \emph{Statistical Science}, \textbf{20}(1), 50--67.

\bibitem[{Kamishima(2003)}]{Kamishima2003}
Kamishima T (2003).
\newblock \enquote{Nantonac Collaborative Filtering: Recommendation Based on
  Order Responses.}
\newblock In \emph{Proceedings of the Ninth ACM SIGKDD International Conference
  on Knowledge Discovery and Data Mining}, KDD '03, pp. 583--588. ACM, New
  York, NY, USA.
\newblock ISBN 1-58113-737-0.
\newblock \doi{10.1145/956750.956823}.
\newblock \urlprefix\url{http://doi.acm.org/10.1145/956750.956823}.

\bibitem[{Lee and Yu(2013)}]{Lee2013}
Lee PH, Yu PL (2013).
\newblock \enquote{An \proglang{R} Package for Analyzing and Modeling Ranking
  Data.}
\newblock \emph{BMC Medical Research Methodology}, \textbf{13}(1), 65.
\newblock ISSN 1471-2288.
\newblock \doi{10.1186/1471-2288-13-65}.
\newblock \urlprefix\url{https://doi.org/10.1186/1471-2288-13-65}.

\bibitem[{Liu \emph{et~al.}(2019)Liu, Crispino, Scheel, Vitelli, and
  Frigessi}]{Liu2019}
Liu Q, Crispino M, Scheel I, Vitelli V, Frigessi A (2019).
\newblock \enquote{Model-Based Learning from Preference Data.}
\newblock \emph{Annual Review of Statistics and Its Application},
  \textbf{6}(1).
\newblock \doi{10.1146/annurev-statistics-031017-100213}.

\bibitem[{Lu and Boutilier(2014)}]{Lu2015}
Lu T, Boutilier C (2014).
\newblock \enquote{Effective Sampling and Learning for {M}allows Models with
  Pairwise-Preference Data.}
\newblock \emph{Journal of Machine Learning Research}, \textbf{15}, 3783--3829.

\bibitem[{Mallows(1957)}]{Mallows1957}
Mallows CL (1957).
\newblock \enquote{Non-Null Ranking Models. I.}
\newblock \emph{Biometrika}, \textbf{44}(1-2), 114--130.
\newblock \doi{10.1093/biomet/44.1-2.114}.
\newblock \urlprefix\url{http://dx.doi.org/10.1093/biomet/44.1-2.114}.

\bibitem[{Marden(1995)}]{Marden1995}
Marden JI (1995).
\newblock \emph{Analyzing and Modeling Rank Data}, volume~64 of
  \emph{Monographs on Statistics and Applied Probability}.
\newblock Chapman \& Hall, Cambridge, MA, USA.

\bibitem[{Marquis~of Condorcet(1785)}]{Condorcet1785}
Marquis~of Condorcet MJANdC (1785).
\newblock \enquote{Essai sur l'application de l'analyse {\`a} la
  probabilit{\'e} des d{\'e}cisions rendues {\`a} la pluralit{\'e} des voix.}
\newblock \emph{Paris: De l'imprimerie royale}.

\bibitem[{Meil\v{a} and Chen(2010)}]{Meila2010}
Meil\v{a} M, Chen H (2010).
\newblock \enquote{{D}irichlet Process Mixtures of Generalized {M}allows
  Models.}
\newblock In \emph{Proceedings of the Twenty-Sixth Conference Annual Conference
  on Uncertainty in Artificial Intelligence (UAI-10)}, pp. 358--367. AUAI
  Press, Corvallis, OR, USA.

\bibitem[{Meredith and Kruschke(2018)}]{Meredith2018}
Meredith M, Kruschke J (2018).
\newblock \emph{\pkg{HDInterval}: Highest (Posterior) Density Intervals}.
\newblock \proglang{R} package version 0.2.0,
  \urlprefix\url{https://CRAN.R-project.org/package=HDInterval}.

\bibitem[{Mersmann(2018)}]{Mersmann2018}
Mersmann O (2018).
\newblock \emph{\pkg{microbenchmark}: Accurate Timing Functions}.
\newblock R package version 1.4-6,
  \urlprefix\url{https://CRAN.R-project.org/package=microbenchmark}.

\bibitem[{Meyer and Hornik(2009)}]{Meyer2009}
Meyer D, Hornik K (2009).
\newblock \enquote{Generalized and Customizable Sets in \proglang{R}.}
\newblock \emph{Journal of Statistical Software}, \textbf{31}(2), 1--27.
\newblock \doi{10.18637/jss.v031.i02}.

\bibitem[{Meyer and Hornik(2018)}]{Meyer2018}
Meyer D, Hornik K (2018).
\newblock \emph{\pkg{relations}: Data Structures and Algorithms for Relations}.
\newblock \proglang{R} package version 0.6-8,
  \urlprefix\url{https://CRAN.R-project.org/package=relations}.

\bibitem[{Mukherjee(2016)}]{Mukherjee2016}
Mukherjee S (2016).
\newblock \enquote{Estimation in Exponential Families on Permutations.}
\newblock \emph{The Annals of Statistics}, \textbf{44}(2), 853--875.

\bibitem[{M{\"u}ller and Wickham(2018)}]{Muller2018}
M{\"u}ller K, Wickham H (2018).
\newblock \emph{\pkg{tibble}: Simple Data Frames}.
\newblock \proglang{R} package version 1.4.2,
  \urlprefix\url{https://CRAN.R-project.org/package=tibble}.

\bibitem[{Papastamoulis(2016)}]{Papastamoulis2015}
Papastamoulis P (2016).
\newblock \enquote{\pkg{label.switching}: An \proglang{R} Package for Dealing
  with the Label Switching Problem in {MCMC} Outputs.}
\newblock \emph{Journal of Statistical Software}, \textbf{69}.

\bibitem[{Qian(2018)}]{rankdist}
Qian Z (2018).
\newblock \emph{\pkg{rankdist}: Distance Based Ranking Models}.
\newblock \proglang{R} package version 1.1-3,
  \urlprefix\url{https://CRAN.R-project.org/package=rankdist}.

\bibitem[{{\proglang{R} Core Team}(2018{\natexlab{a}})}]{R}
{\proglang{R} Core Team} (2018{\natexlab{a}}).
\newblock \emph{\proglang{R}: A Language and Environment for Statistical
  Computing}.
\newblock \proglang{R} Foundation for Statistical Computing, Vienna, Austria.
\newblock \urlprefix\url{https://www.R-project.org/}.

\bibitem[{{\proglang{R} Core Team}(2018{\natexlab{b}})}]{RExt}
{\proglang{R} Core Team} (2018{\natexlab{b}}).
\newblock \emph{Writing \proglang{R} Extensions}.
\newblock \proglang{R} Foundation for Statistical Computing, Vienna, Austria.
\newblock
  \urlprefix\url{https://cran.r-project.org/doc/manuals/r-release/R-exts.html}.

\bibitem[{Sanderson and Curtin(2016)}]{Sanderson2016}
Sanderson C, Curtin R (2016).
\newblock \enquote{\pkg{Armadillo}: a template-based \proglang{C++} library for
  linear algebra.}
\newblock \emph{Journal of Open Source Software}, \textbf{1}.

\bibitem[{Sanderson and Curtin(2018)}]{Sanderson2018}
Sanderson C, Curtin R (2018).
\newblock \enquote{A User-Friendly Hybrid Sparse Matrix Class in
  \proglang{C++}.}
\newblock In JH~Davenport, M~Kauers, G~Labahn, J~Urban (eds.),
  \emph{Mathematical Software -- ICMS 2018}, pp. 422--430. Springer
  International Publishing.

\bibitem[{Sloane(2017)}]{Sloane2017}
Sloane NJA (2017).
\newblock \enquote{The {O}n-{L}ine {E}ncyclopedia of {I}nteger {S}equences.}
\newblock \urlprefix\url{http://oeis.org}.

\bibitem[{Stephens(2000)}]{Stephens2000}
Stephens M (2000).
\newblock \enquote{Dealing with label switching in mixture models.}
\newblock \emph{Journal of the Royal Statistical Society B}, \textbf{62}(4),
  795--809.
\newblock \doi{10.1111/1467-9868.00265}.
\newblock
  \urlprefix\url{https://rss.onlinelibrary.wiley.com/doi/abs/10.1111/1467-9868.00265}.

\bibitem[{Vitelli \emph{et~al.}(2018)Vitelli, S{\o}rensen, Crispino, Frigessi,
  and Arjas}]{Vitelli2018}
Vitelli V, S{\o}rensen {\O}, Crispino M, Frigessi A, Arjas E (2018).
\newblock \enquote{Probabilistic Preference Learning with the Mallows Rank
  Model.}
\newblock \emph{Journal of Machine Learning Research}, \textbf{18}(1),
  5796--5844.
\newblock ISSN 1532-4435.

\bibitem[{Wickham(2016)}]{Wickham2016}
Wickham H (2016).
\newblock \emph{\pkg{ggplot2}: Elegant Graphics for Data Analysis}.
\newblock Springer-Verlag New York.
\newblock ISBN 978-3-319-24277-4.
\newblock \urlprefix\url{http://ggplot2.org}.

\bibitem[{Wickham \emph{et~al.}(2018)Wickham, Fran{\c c}ois, Henry, and
  M{\"u}ller}]{Wickham2018dplyr}
Wickham H, Fran{\c c}ois R, Henry L, M{\"u}ller K (2018).
\newblock \emph{\pkg{dplyr}: A Grammar of Data Manipulation}.
\newblock \proglang{R} package version 0.7.7,
  \urlprefix\url{https://CRAN.R-project.org/package=dplyr}.

\bibitem[{Wickham and Henry(2018)}]{Wickham2018tidyr}
Wickham H, Henry L (2018).
\newblock \emph{\pkg{tidyr}: Easily Tidy Data with 'spread()' and 'gather()'
  Functions}.
\newblock \proglang{R} package version 0.8.1,
  \urlprefix\url{https://CRAN.R-project.org/package=tidyr}.

\bibitem[{Wilke(2018)}]{Wilke2018}
Wilke CO (2018).
\newblock \emph{\pkg{cowplot}: Streamlined Plot Theme and Plot Annotations for
  'ggplot2'}.
\newblock \proglang{R} package version 0.9.3,
  \urlprefix\url{https://CRAN.R-project.org/package=cowplot}.

\end{thebibliography}

\end{document}